
\font\eightrm=cmr8 at 8pt

\font\seventeenrm=cmr17 at 17pt
\font\twentyonerm=cmr17 at 21pt

\font\ss=cmss10

\font\csc=cmcsc10

\font\twelvecal=cmsy10 at 12pt

\font\twelvemath=cmmi12

\font\seventeenbold=cmbx7 at 17pt

\font\fively=lasy5
\font\sevenly=lasy7
\font\tenly=lasy10

\textfont10=\tenly
\scriptfont10=\sevenly
\scriptscriptfont10=\fively
\magnification=1200
\parskip=10pt
\parindent=20pt
\def\today{\ifcase\month\or January\or February\or March\or April\or May\or
June
       \or July\or August\or September\or October\or November\or December\fi
       \space\number\day, \number\year}

\def\title#1{\footline={\ifnum\pageno<2\hfil
       \else\hss\tenrm\folio\hss\fi}\vskip1truein\centerline{{#1}
       \footnote{\raise1ex\hbox{*}}{\eightrm Supported in part
       by the Robert A. Welch Foundation and N.S.F. Grants
       PHY-880637 and\break PHY-8605978.}}}

\def\newpage{\vfill\eject}
\def\abstract#1{\centerline{\bf ABSTRACT}\vskip.2truein{\narrower\noindent#1
       \smallskip}}

\def\runninghead#1#2{\voffset=2\baselineskip\nopagenumbers
       \headline={\ifodd\pageno\rightheadline\else \leftheadline\fi}
       \def\rightheadline{{\sl#1}\hfill{\rm\folio}}
       \def\leftheadline{{\rm\folio}\hfill{\sl#2}}}
\def\SS{\mathhexbox278}

\newcount\footnoteno
\def\Footnote#1{\advance\footnoteno by 1
                \let\SF=\empty
                \ifhmode\edef\SF{\spacefactor=\the\spacefactor}\/\fi
                $^{\the\footnoteno}$\ignorespaces
                \SF\vfootnote{$^{\the\footnoteno}$}{#1}}

\def\place#1#2#3{\vbox to0pt{\kern-\parskip\kern-7pt
                             \kern-#2truein\hbox{\kern#1truein #3}
                             \vss}\nointerlineskip}
\def\figurecaption#1#2{\kern.75truein\vbox{\hsize=5truein\noindent{\bf Figure
    \figlabel{#1}:} #2}}
\def\tablecaption#1#2{\kern.75truein\lower12truept\hbox{\vbox{\hsize=5truein
    \noindent{\bf Table\hskip5truept\tablabel{#1}:} #2}}}
\def\boxed#1{\lower3pt\hbox{
                       \vbox{\hrule\hbox{\vrule

\vbox{\kern2pt\hbox{\kern3pt#1\kern3pt}\kern3pt}\vrule}
                         \hrule}}}

\def\g{\gamma}
\def\D{\Delta}

\def\k{\kappa}
\def\l{\lambda}\def\L{\Lambda}
\def\m{\mu}
\def\n{\nu}

\def\p{\pi}\def\P{\Pi}

\def\t{\tau}

\def\ca#1{\relax\ifmmode {{\cal #1}}\else $\cal #1$\fi}

\def\calb{{\cal B}}

\def\calm{{\cal M}}

\def\inbar{\vrule height1.5ex width.4pt depth0pt}
\def\IB{\relax{\rm I\kern-.18em B}}
\def\IC{\relax\hbox{\kern.25em$\inbar\kern-.3em{\rm C}$}}
\def\ID{\relax{\rm I\kern-.18em D}}
\def\IE{\relax{\rm I\kern-.18em E}}
\def\IF{\relax{\rm I\kern-.18em F}}
\def\IG{\relax\hbox{\kern.25em$\inbar\kern-.3em{\rm G}$}}
\def\IH{\relax{\rm I\kern-.18em H}}
\def\II{\relax{\rm I\kern-.18em I}}
\def\IK{\relax{\rm I\kern-.18em K}}
\def\IL{\relax{\rm I\kern-.18em L}}
\def\IM{\relax{\rm I\kern-.18em M}}
\def\IN{\relax{\rm I\kern-.18em N}}
\def\IO{\relax\hbox{\kern.25em$\inbar\kern-.3em{\rm O}$}}
\def\IP{\relax{\rm I\kern-.18em P}}
\def\IQ{\relax\hbox{\kern.25em$\inbar\kern-.3em{\rm Q}$}}
\def\IR{\relax{\rm I\kern-.18em R}}
\def\IZ{\relax\ifmmode\hbox{\ss Z\kern-.4em Z}\else{\ss Z\kern-.4em Z}\fi}
\def\IGa{\relax{\rm I}\kern-.18em\Gamma}
\def\IPi{\relax{\rm I}\kern-.18em\Pi}
\def\ITh{\relax\hbox{\kern.25em$\inbar\kern-.3em\Theta$}}
\def\IOm{\relax\thinspace\inbar\kern1.95pt\inbar\kern-5.525pt\Omega}


\def\ie{{\it i.e.,\ \/}}
\def\eg{{\it e.g.,\ \/}}
\def\noblackboxes{\overfullrule=0pt}

\def\cy{Calabi--Yau}
\def\cym{Calabi--Yau manifold}
\def\cys{Calabi--Yau manifolds}
\def\cyt{Calabi--Yau threefold}

\def\K{K\"ahler}

\def\H#1#2{\relax\ifmmode {H^{#1#2}}\else $H^{#1 #2}$\fi}
\def\M{\relax\ifmmode{\calm}\else $\calm$\fi}

\def\Bigcheck{\lower3.8pt\hbox{\smash{\hbox{{\twentyonerm \v{}}}}}}
\def\bigboldcheck{\smash{\hbox{{\seventeenbold\v{}}}}}

\def\Bighat{\lower3.8pt\hbox{\smash{\hbox{{\twentyonerm \^{}}}}}}

\def\Msharp{\relax\ifmmode{\calm^\sharp}\else $\smash{\calm^\sharp}$\fi}
\def\Mflat{\relax\ifmmode{\calm^\flat}\else $\smash{\calm^\flat}$\fi}
\def\preMcheck{\kern2pt\hbox{\Bigcheck\kern-12pt{$\cal M$}}}
\def\Mcheck{\relax\ifmmode\preMcheck\else $\preMcheck$\fi}
\def\preMhat{\kern2pt\hbox{\Bighat\kern-12pt{$\cal M$}}}
\def\Mhat{\relax\ifmmode\preMhat\else $\preMhat$\fi}

\def\Bsharp{\relax\ifmmode{\calb^\sharp}\else $\calb^\sharp$\fi}
\def\Bflat{\relax\ifmmode{\calb^\flat}\else $\calb^\flat$ \fi}
\def\preBcheck{\hbox{\Bigcheck\kern-9pt{$\cal B$}}}
\def\Bcheck{\relax\ifmmode\preBcheck\else $\preBcheck$\fi}
\def\preBhat{\hbox{\Bighat\kern-9pt{$\cal B$}}}
\def\Bhat{\relax\ifmmode\preBhat\else $\preBhat$\fi}

\def\figBcheck{\kern3pt\hbox{\raise1pt\hbox{\bigboldcheck}\kern-11pt
    {\twelvecal B}}}
\def\figBsharp{{\twelvecal B}\raise5pt\hbox{$\twelvemath\sharp$}}
\def\figBflat{{\twelvecal B}\raise5pt\hbox{$\twelvemath\flat$}}

\def\gcheck{\hbox{\lower2.5pt\hbox{\Bigcheck}\kern-8pt$\g$}}
\def\lhat{\hbox{\raise.5pt\hbox{\Bighat}\kern-8pt$\l$}}

\def\Fcheck{\kern2pt\hbox{\raise1pt\hbox{\Bigcheck}\kern-10pt{$\cal F$}}}
\def\Fhat{\kern2pt\hbox{\raise1pt\hbox{\Bighat}\kern-10pt{$\cal F$}}}

\def\cp#1{\relax\ifmmode {\IP\kern-2pt{}_{#1}}\else $\IP\kern-2pt{}_{#1}$\fi}
\def\h#1#2{\relax\ifmmode {b_{#1#2}}\else $b_{#1#2}$\fi}

\def\frac#1#2{{#1\over #2}}

\def\cone{\relax\thinspace\hbox{$<\kern-.8em{)}$}}
\mathchardef\mho"0A30

\def\-{\hphantom{-}}



\def\picture #1 by #2 (#3){\vbox to #2{\hrule width #1 height 0pt depth 0pt
                                       \vfill\special{picture #3}}}
\def\scaledpicture #1 by #2 (#3 scaled #4){{\dimen0=#1 \dimen1=#2
           \divide\dimen0 by 1000 \multiply\dimen0 by #4
            \divide\dimen1 by 1000 \multiply\dimen1 by #4
            \picture \dimen0 by \dimen1 (#3 scaled #4)}}
\def\illustration #1 by #2 (#3){\vbox to #2{\hrule width #1 height 0pt depth
0pt
                                       \vfill\special{illustration #3}}}
\def\scaledillustration #1 by #2 (#3 scaled #4){{\dimen0=#1 \dimen1=#2
           \divide\dimen0 by 1000 \multiply\dimen0 by #4
            \divide\dimen1 by 1000 \multiply\dimen1 by #4
            \illustration \dimen0 by \dimen1 (#3 scaled #4)}}


\def\delaOssa{\nobreak\vskip1truein\hbox to\hsize
       {\hskip 4truein Xenia de la Ossa\hfill}}

\def\hoy{\number\day\space de \ifcase\month\or enero\or febrero\or marzo\or
       abril\or mayo\or junio\or julio\or agosto\or septiembre\or octubre\or
       noviembre\or diciembre\fi\space de \number\year}

\def\cropen#1{\crcr\noalign{\vskip #1}}

\newif\ifproofmode
\proofmodefalse

\newif\ifforwardreference
\forwardreferencefalse

\newif\ifchapternumbers
\chapternumbersfalse

\newif\ifcontinuousnumbering
\continuousnumberingfalse

\newif\iffigurechapternumbers
\figurechapternumbersfalse

\newif\ifcontinuousfigurenumbering
\continuousfigurenumberingfalse

\newif\iftablechapternumbers
\tablechapternumbersfalse

\newif\ifcontinuoustablenumbering
\continuoustablenumberingfalse

\font\eqsixrm=cmr6

\def\marginstyle{\eqsixrm}

\newtoks\chapletter
\newcount\chapno
\newcount\eqlabelno
\newcount\figureno
\newcount\tableno

\chapno=0
\eqlabelno=0
\figureno=0
\tableno=0

\def\chapfolio{\ifnum\chapno>0 \the\chapno\else\the\chapletter\fi}

\def\bumpchapno{\ifnum\chapno>-1 \global\advance\chapno by 1
\else\global\advance\chapno by -1 \setletter\chapno\fi
\ifcontinuousnumbering\else\global\eqlabelno=0 \fi
\ifcontinuousfigurenumbering\else\global\figureno=0 \fi
\ifcontinuoustablenumbering\else\global\tableno=0 \fi}

\def\setletter#1{\ifcase-#1{}\or{}%
\or\global\chapletter={A}%
\or\global\chapletter={B}%
\or\global\chapletter={C}%
\or\global\chapletter={D}%
\or\global\chapletter={E}%
\or\global\chapletter={F}%
\or\global\chapletter={G}%
\or\global\chapletter={H}%
\or\global\chapletter={I}%
\or\global\chapletter={J}%
\or\global\chapletter={K}%
\or\global\chapletter={L}%
\or\global\chapletter={M}%
\or\global\chapletter={N}%
\or\global\chapletter={O}%
\or\global\chapletter={P}%
\or\global\chapletter={Q}%
\or\global\chapletter={R}%
\or\global\chapletter={S}%
\or\global\chapletter={T}%
\or\global\chapletter={U}%
\or\global\chapletter={V}%
\or\global\chapletter={W}%
\or\global\chapletter={X}%
\or\global\chapletter={Y}%
\or\global\chapletter={Z}\fi}

\def\tempsetletter#1{\ifcase-#1{}\or{}%
\or\global\chapletter={A}%
\or\global\chapletter={B}%
\or\global\chapletter={C}%
\or\global\chapletter={D}%
\or\global\chapletter={E}%
\or\global\chapletter={F}%
\or\global\chapletter={G}%
\or\global\chapletter={H}%
\or\global\chapletter={I}%
\or\global\chapletter={J}%
\or\global\chapletter={K}%
\or\global\chapletter={L}%
\or\global\chapletter={M}%
\or\global\chapletter={N}%
\or\global\chapletter={O}%
\or\global\chapletter={P}%
\or\global\chapletter={Q}%
\or\global\chapletter={R}%
\or\global\chapletter={S}%
\or\global\chapletter={T}%
\or\global\chapletter={U}%
\or\global\chapletter={V}%
\or\global\chapletter={W}%
\or\global\chapletter={X}%
\or\global\chapletter={Y}%
\or\global\chapletter={Z}\fi}

\def\chapshow#1{\ifnum#1>0 \relax#1%
\else{\tempsetletter{\number#1}\chapno=#1\chapfolio}\fi}

\def\initialeqmacro{\ifproofmode
\immediate\openout2=allcrossreferfile \fi
\ifforwardreference\expandafter \def \csname CHAPLABELintro\endcsname {1}
\expandafter \def \csname EQLABEL8dim\endcsname {1.1?}
\expandafter \def \csname EQLABEL6dim\endcsname {1.2?}
\expandafter \def \csname EQLABEL4dim\endcsname {1.3?}
\expandafter \def \csname CHAPLABELtoric\endcsname {2}
\expandafter \def \csname EQLABELhodgethree\endcsname {2.1?}
\expandafter \def \csname EQLABELhodgefour\endcsname {2.2?}
\expandafter \def \csname EQLABELWei\endcsname {2.3?}
\expandafter \def \csname TABLABELscaling\endcsname {2.1?}
\expandafter \def \csname TABLABELscalingtwo\endcsname {2.2?}
\expandafter \def \csname CHAPLABELbundles\endcsname {3}
\expandafter \def \csname EQLABELNbundle\endcsname {3.1?}
\expandafter \def \csname EQLABELzeroN\endcsname {3.2?}
\expandafter \def \csname EQLABELchernN\endcsname {3.3?}
\expandafter \def \csname EQLABELgamma\endcsname {3.4?}
\expandafter \def \csname EQLABELchernV\endcsname {3.5?}
\expandafter \def \csname EQLABELgammasq\endcsname {3.6?}
\expandafter \def \csname EQLABELchernE\endcsname {3.7?}
\expandafter \def \csname EQLABELtadpole\endcsname {3.8?}
\expandafter \def \csname EQLABELhetanomaly\endcsname {3.9?}
\expandafter \def \csname EQLABELbundlehodge\endcsname {3.10?}
\expandafter \def \csname EQLABELIdef\endcsname {3.11?}
\expandafter \def \csname EQLABELItau\endcsname {3.12?}
\expandafter \def \csname EQLABELISUN\endcsname {3.13?}
\expandafter \def \csname EQLABELTodd\endcsname {3.14?}
\expandafter \def \csname EQLABELh11old\endcsname {3.15?}
\expandafter \def \csname EQLABELh11new\endcsname {3.16?}
\expandafter \def \csname CHAPLABELmap\endcsname {4}
\expandafter \def \csname EQLABELtdef\endcsname {4.1?}
\expandafter \def \csname EQLABELetas\endcsname {4.2?}
\expandafter \def \csname TABLABELfmnp\endcsname {4.1?}
\expandafter \def \csname EQLABELfmnprels\endcsname {4.3?}
\expandafter \def \csname EQLABELetamin\endcsname {4.4?}
\expandafter \def \csname CHAPLABELfivebranes\endcsname {5}
\expandafter \def \csname CHAPLABELfin\endcsname {6}

\ifproofmode\immediate\openout1=labelfile \fi\fi}

\def\initialeqmacros{\ifproofmode
\immediate\openout2=allcrossreferfile \fi
\ifforwardreference\expandafter \def \csname CHAPLABELintro\endcsname {1}
\expandafter \def \csname EQLABEL8dim\endcsname {1.1?}
\expandafter \def \csname EQLABEL6dim\endcsname {1.2?}
\expandafter \def \csname EQLABEL4dim\endcsname {1.3?}
\expandafter \def \csname CHAPLABELtoric\endcsname {2}
\expandafter \def \csname EQLABELhodgethree\endcsname {2.1?}
\expandafter \def \csname EQLABELhodgefour\endcsname {2.2?}
\expandafter \def \csname EQLABELWei\endcsname {2.3?}
\expandafter \def \csname TABLABELscaling\endcsname {2.1?}
\expandafter \def \csname TABLABELscalingtwo\endcsname {2.2?}
\expandafter \def \csname CHAPLABELbundles\endcsname {3}
\expandafter \def \csname EQLABELNbundle\endcsname {3.1?}
\expandafter \def \csname EQLABELzeroN\endcsname {3.2?}
\expandafter \def \csname EQLABELchernN\endcsname {3.3?}
\expandafter \def \csname EQLABELgamma\endcsname {3.4?}
\expandafter \def \csname EQLABELchernV\endcsname {3.5?}
\expandafter \def \csname EQLABELgammasq\endcsname {3.6?}
\expandafter \def \csname EQLABELchernE\endcsname {3.7?}
\expandafter \def \csname EQLABELtadpole\endcsname {3.8?}
\expandafter \def \csname EQLABELhetanomaly\endcsname {3.9?}
\expandafter \def \csname EQLABELbundlehodge\endcsname {3.10?}
\expandafter \def \csname EQLABELIdef\endcsname {3.11?}
\expandafter \def \csname EQLABELItau\endcsname {3.12?}
\expandafter \def \csname EQLABELISUN\endcsname {3.13?}
\expandafter \def \csname EQLABELTodd\endcsname {3.14?}
\expandafter \def \csname EQLABELh11old\endcsname {3.15?}
\expandafter \def \csname EQLABELh11new\endcsname {3.16?}
\expandafter \def \csname CHAPLABELmap\endcsname {4}
\expandafter \def \csname EQLABELtdef\endcsname {4.1?}
\expandafter \def \csname EQLABELetas\endcsname {4.2?}
\expandafter \def \csname TABLABELfmnp\endcsname {4.1?}
\expandafter \def \csname EQLABELfmnprels\endcsname {4.3?}
\expandafter \def \csname EQLABELetamin\endcsname {4.4?}
\expandafter \def \csname CHAPLABELfivebranes\endcsname {5}
\expandafter \def \csname CHAPLABELfin\endcsname {6}

\ifproofmode\immediate\openout1=labelfile \fi\fi}

\def\chaplabel#1{\bumpchapno\ifproofmode\ifforwardreference
\immediate\write1{\noexpand\expandafter\noexpand\def
\noexpand\csname CHAPLABEL#1\endcsname{\the\chapno}}\fi\fi
\global\expandafter\edef\csname CHAPLABEL#1\endcsname
{\the\chapno}\ifproofmode\llap{\hbox{\marginstyle #1\ }}\fi\chapfolio}

\def\eqnum{\global\advance\eqlabelno by 1
\eqno(\ifchapternumbers\chapfolio.\fi\the\eqlabelno)}

\def\eqlabel#1{\global\advance\eqlabelno by 1 \ifproofmode\ifforwardreference
\immediate\write1{\noexpand\expandafter\noexpand\def
\noexpand\csname EQLABEL#1\endcsname{\the\chapno.\the\eqlabelno?}}\fi\fi
\global\expandafter\edef\csname EQLABEL#1\endcsname
{\the\chapno.\the\eqlabelno?}\eqno(\ifchapternumbers\chapfolio.\fi
\the\eqlabelno)\ifproofmode\rlap{\hbox{\marginstyle #1}}\fi}

\def\eqalignnum{\global\advance\eqlabelno by 1
&(\ifchapternumbers\chapfolio.\fi\the\eqlabelno)}

\def\eqalignlabel#1{\global\advance\eqlabelno by 1 \ifproofmode
\ifforwardreference\immediate\write1{\noexpand\expandafter\noexpand\def
\noexpand\csname EQLABEL#1\endcsname{\the\chapno.\the\eqlabelno?}}\fi\fi
\global\expandafter\edef\csname EQLABEL#1\endcsname
{\the\chapno.\the\eqlabelno?}&(\ifchapternumbers\chapfolio.\fi
\the\eqlabelno)\ifproofmode\rlap{\hbox{\marginstyle #1}}\fi}

\def\eqref#1{\hbox{(\ifundefined{EQLABEL#1}***)\ifproofmode\ifforwardreference%
\else\write16{ ***Undefined Equation Reference #1*** }\fi
\else\write16{ ***Undefined Equation Reference #1*** }\fi
\else\edef\LABxx{\getlabel{EQLABEL#1}}%
\def\LAByy{\expandafter\stripchap\LABxx}\ifchapternumbers%
\chapshow{\LAByy}.\expandafter\stripeq\LABxx%
\else\ifnum\number\LAByy=\chapno\relax\expandafter\stripeq\LABxx%
\else\chapshow{\LAByy}.\expandafter\stripeq\LABxx\fi\fi)\fi}%
\ifproofmode\write2{Equation #1}\fi}

\def\fignum{\global\advance\figureno by 1
\relax\iffigurechapternumbers\chapfolio.\fi\the\figureno}

\def\figlabel#1{\global\advance\figureno by 1
\relax\ifproofmode\ifforwardreference
\immediate\write1{\noexpand\expandafter\noexpand\def
\noexpand\csname FIGLABEL#1\endcsname{\the\chapno.\the\figureno?}}\fi\fi
\global\expandafter\edef\csname FIGLABEL#1\endcsname
{\the\chapno.\the\figureno?}\iffigurechapternumbers\chapfolio.\fi
\ifproofmode\llap{\hbox{\marginstyle#1
\kern1.2truein}}\relax\fi\the\figureno}

\def\figref#1{\hbox{\ifundefined{FIGLABEL#1}!!!!\ifproofmode\ifforwardreference%
\else\write16{ ***Undefined Figure Reference #1*** }\fi
\else\write16{ ***Undefined Figure Reference #1*** }\fi
\else\edef\LABxx{\getlabel{FIGLABEL#1}}%
\def\LAByy{\expandafter\stripchap\LABxx}\iffigurechapternumbers%
\chapshow{\LAByy}.\expandafter\stripeq\LABxx%
\else\ifnum \number\LAByy=\chapno\relax\expandafter\stripeq\LABxx%
\else\chapshow{\LAByy}.\expandafter\stripeq\LABxx\fi\fi\fi}%
\ifproofmode\write2{Figure #1}\fi}

\def\tabnum{\global\advance\tableno by 1
\relax\iftablechapternumbers\chapfolio.\fi\the\tableno}

\def\tablabel#1{\global\advance\tableno by 1
\relax\ifproofmode\ifforwardreference
\immediate\write1{\noexpand\expandafter\noexpand\def
\noexpand\csname TABLABEL#1\endcsname{\the\chapno.\the\tableno?}}\fi\fi
\global\expandafter\edef\csname TABLABEL#1\endcsname
{\the\chapno.\the\tableno?}\iftablechapternumbers\chapfolio.\fi
\ifproofmode\llap{\hbox{\marginstyle#1
\kern1.2truein}}\relax\fi\the\tableno}

\def\tabref#1{\hbox{\ifundefined{TABLABEL#1}!!!!\ifproofmode\ifforwardreference%
\else\write16{ ***Undefined Table Reference #1*** }\fi
\else\write16{ ***Undefined Table Reference #1*** }\fi
\else\edef\LABtt{\getlabel{TABLABEL#1}}%
\def\LABTT{\expandafter\stripchap\LABtt}\iftablechapternumbers%
\chapshow{\LABTT}.\expandafter\stripeq\LABtt%
\else\ifnum\number\LABTT=\chapno\relax\expandafter\stripeq\LABtt%
\else\chapshow{\LABTT}.\expandafter\stripeq\LABtt\fi\fi\fi}%
\ifproofmode\write2{Table#1}\fi}

\newdimen\sectionskip     \sectionskip=20truept
\newcount\sectno
\def\section#1#2{\sectno=0 \null\vskip\sectionskip
    \centerline{\chaplabel{#1}.~~{\bf#2}}\nobreak\vskip.2truein
    \noindent\ignorespaces}

\def\advancesectno{\global\advance\sectno by 1}
\def\sectfolio{\number\sectno}
\def\subsection#1{\goodbreak\advancesectno\null\vskip10pt
                  \noindent\chapfolio.~\sectfolio.~{\bf #1}
                  \nobreak\vskip.05truein\noindent\ignorespaces}

\def\uttg#1{\null\vskip.1truein
    \ifproofmode \line{\hfill{\bf Draft}:
    UTTG--{#1}--\number\year}\line{\hfill\today}
    \else \line{\hfill UTTG--{#1}--\number\year}
    \line{\hfill\ifcase\month\or January\or February\or March\or April\or
May\or June
    \or July\or August\or September\or October\or November\or December\fi
    \space\number\year}\fi}

\def\getlabel#1{\csname#1\endcsname}
\def\ifundefined#1{\expandafter\ifx\csname#1\endcsname\relax}
\def\stripchap#1.#2?{#1}
\def\stripeq#1.#2?{#2}

%
\catcode`@=11 
\def\space@ver#1{\let\@sf=\empty\ifmmode#1\else\ifhmode%
\edef\@sf{\spacefactor=\the\spacefactor}\unskip${}#1$\relax\fi\fi}
\newcount\referencecount     \referencecount=0
\newif\ifreferenceopen       \newwrite\referencewrite
\newtoks\rw@toks
\def\refmark#1{\relax[#1]}
\def\refend{\refmark{\number\referencecount}}
\newcount\lastrefsbegincount \lastrefsbegincount=0
\def\refsend{\refmark{\count255=\referencecount%
\advance\count255 by -\lastrefsbegincount%
\ifcase\count255 \number\referencecount%
\or\number\lastrefsbegincount,\number\referencecount%
\else\number\lastrefsbegincount-\number\referencecount\fi}}
\def\refch@ck{\chardef\rw@write=\referencewrite
\ifreferenceopen\else\referenceopentrue
\immediate\openout\referencewrite=referenc.texauxil \fi}
%
{\catcode`\^^M=\active 
  \gdef\obeyendofline{\catcode`\^^M\active \let^^M\ }}%
%
{\catcode`\^^M=\active 
  \gdef\ignoreendofline{\catcode`\^^M=5}}
{\obeyendofline\gdef\rw@start#1{\def\t@st{#1}\ifx\t@st\blankend%
\endgroup\@sf\relax\else\ifx\t@st\bl@nkend\endgroup\@sf\relax%
\else\rw@begin#1
\backtotext
\fi\fi}}
{\obeyendofline\gdef\rw@begin#1
{\def\n@xt{#1}\rw@toks={#1}\relax%
\rw@next}}
\def\blankend{}
{\obeylines\gdef\bl@nkend{
}}
\newif\iffirstrefline  \firstreflinetrue
\def\rwr@teswitch{\ifx\n@xt\blankend\let\n@xt=\rw@begin%
\else\iffirstrefline\global\firstreflinefalse%
\immediate\write\rw@write{\noexpand\obeyendofline\the\rw@toks}%
\let\n@xt=\rw@begin%
\else\ifx\n@xt\rw@@d \def\n@xt{\immediate\write\rw@write{%
\noexpand\ignoreendofline}\endgroup\@sf}%
\else\immediate\write\rw@write{\the\rw@toks}%
\let\n@xt=\rw@begin\fi\fi\fi}
\def\rw@next{\rwr@teswitch\n@xt}
\def\rw@@d{\backtotext} \let\rw@end=\relax
\let\backtotext=\relax

\newdimen\refindent     \refindent=30pt
\def\Textindent#1{\noindent\llap{#1\enspace}\ignorespaces}
\def\refitem#1{\par\hangafter=0 \hangindent=\refindent\Textindent{#1}}
\def\REFNUM#1{\space@ver{}\refch@ck\firstreflinetrue%
\global\advance\referencecount by 1 \xdef#1{\the\referencecount}}
\def\refnum#1{\space@ver{}\refch@ck\firstreflinetrue%
\global\advance\referencecount by 1\xdef#1{\the\referencecount}\refend}

\def\REF#1{\REFNUM#1%
\immediate\write\referencewrite{%
\noexpand\refitem{#1.}}%
\begingroup\obeyendofline\rw@start}
\def\ref{\refnum\?%
\immediate\write\referencewrite{\noexpand\refitem{\?.}}%
\begingroup\obeyendofline\rw@start}
\def\Ref#1{\refnum#1%
\immediate\write\referencewrite{\noexpand\refitem{#1.}}%
\begingroup\obeyendofline\rw@start}
\def\REFS#1{\REFNUM#1\global\lastrefsbegincount=\referencecount%
\immediate\write\referencewrite{\noexpand\refitem{#1.}}%
\begingroup\obeyendofline\rw@start}

\def\REFSCON#1{\REF#1}

\def\cite#1{\refmark#1}
\catcode`@=12 
%

\baselineskip=15pt plus 1pt minus 1pt
\parskip=5pt
\chapternumberstrue
\forwardreferencetrue
\figurechapternumberstrue
\tablechapternumberstrue
\ifproofmode
\immediate\openout2=allcrossreferfile \fi
\ifforwardreference\expandafter \def \csname CHAPLABELintro\endcsname {1}
\expandafter \def \csname EQLABEL8dim\endcsname {1.1?}
\expandafter \def \csname EQLABEL6dim\endcsname {1.2?}
\expandafter \def \csname EQLABEL4dim\endcsname {1.3?}
\expandafter \def \csname CHAPLABELtoric\endcsname {2}
\expandafter \def \csname EQLABELhodgethree\endcsname {2.1?}
\expandafter \def \csname EQLABELhodgefour\endcsname {2.2?}
\expandafter \def \csname EQLABELWei\endcsname {2.3?}
\expandafter \def \csname TABLABELscaling\endcsname {2.1?}
\expandafter \def \csname TABLABELscalingtwo\endcsname {2.2?}
\expandafter \def \csname CHAPLABELbundles\endcsname {3}
\expandafter \def \csname EQLABELNbundle\endcsname {3.1?}
\expandafter \def \csname EQLABELzeroN\endcsname {3.2?}
\expandafter \def \csname EQLABELchernN\endcsname {3.3?}
\expandafter \def \csname EQLABELgamma\endcsname {3.4?}
\expandafter \def \csname EQLABELchernV\endcsname {3.5?}
\expandafter \def \csname EQLABELgammasq\endcsname {3.6?}
\expandafter \def \csname EQLABELchernE\endcsname {3.7?}
\expandafter \def \csname EQLABELtadpole\endcsname {3.8?}
\expandafter \def \csname EQLABELhetanomaly\endcsname {3.9?}
\expandafter \def \csname EQLABELbundlehodge\endcsname {3.10?}
\expandafter \def \csname EQLABELIdef\endcsname {3.11?}
\expandafter \def \csname EQLABELItau\endcsname {3.12?}
\expandafter \def \csname EQLABELISUN\endcsname {3.13?}
\expandafter \def \csname EQLABELTodd\endcsname {3.14?}
\expandafter \def \csname EQLABELh11old\endcsname {3.15?}
\expandafter \def \csname EQLABELh11new\endcsname {3.16?}
\expandafter \def \csname CHAPLABELmap\endcsname {4}
\expandafter \def \csname EQLABELtdef\endcsname {4.1?}
\expandafter \def \csname EQLABELetas\endcsname {4.2?}
\expandafter \def \csname TABLABELfmnp\endcsname {4.1?}
\expandafter \def \csname EQLABELfmnprels\endcsname {4.3?}
\expandafter \def \csname EQLABELetamin\endcsname {4.4?}
\expandafter \def \csname CHAPLABELfivebranes\endcsname {5}
\expandafter \def \csname CHAPLABELfin\endcsname {6}

\ifproofmode\immediate\openout1=labelfile \fi\fi
\noblackboxes
\hfuzz=1pt
\vfuzz=2pt


\def\hourandminute{\count255=\time\divide\count255 by 60
\xdef\hour{\number\count255}
\multiply\count255 by -60\advance\count255 by\time
\hour:\ifnum\count255<10 0\fi\the\count255}

\def\subsection#1{\goodbreak\advancesectno\null\vskip10pt
                  \noindent{\it \chapfolio.\sectfolio.~#1}
                  \nobreak\vskip.05truein\noindent\ignorespaces}
\def\cite#1{\refmark{#1}}
\def\\{\hfill\break}
\def\cropen#1{\crcr\noalign{\vskip #1}}

\def\point#1{\noindent\setbox0=\hbox{#1}\kern-\wd0\box0}

\def\cyf{Calabi--Yau fourfold}
\nopagenumbers\pageno=0
\rightline{\eightrm IASSNS-HEP-98/101}\vskip-5pt
\rightline{\eightrm hep-th/9811240}\vskip-5pt

\vskip1truein
\centerline{\seventeenrm Toric Geometry and F-Theory/Heterotic}
\vskip10pt
\centerline{\seventeenrm Duality in Four Dimensions}
\vskip0.5truein
\centerline{\csc Govindan~Rajesh$^{\dag}$}
\vfootnote{$^{\eightrm \dag}$}{\eightrm Email: rajesh@ias.edu.}

\bigskip
\centerline{\it School of Natural Sciences,}
\centerline{\it Institute for Advanced Study,}
\centerline{\it Olden Lane,}
\centerline{\it Princeton, NJ 08540, USA}
\vskip0.75in\bigskip
\nobreak\vbox{
\centerline{\bf ABSTRACT}
\vskip.25truein
\noindent{We study, as hypersurfaces in toric varieties, elliptic \cyf s
for F-theory compactifications dual to  $E_8\times E_8$ heterotic
strings compactified to four dimensions on elliptic \cy\ threefolds with some
choice of vector bundle. We describe how to read off the vector bundle data
for the heterotic compactification from the toric data of the fourfold. This
map allows us to construct, for example, \cy\ fourfolds corresponding to three
generation models with unbroken GUT groups. We also find that the geometry
of the \cyf\ restricts the heterotic
vector bundle data in a manner related to the stability of these bundles.
Finally, we study \cy\ fourfolds
corresponding to heterotic models with fivebranes wrapping curves in the base
of the \cyt s. We find evidence of a topology changing extremal transition on
the fourfold side which corresponds, on the heterotic side, to fivebranes
wrapping different curves in the same homology class in the base.}
} 
\newpage
    {\bf Contents}
\vskip5pt

    1. Introduction 
\vskip3pt

    2. Some results in toric geometry
\vskip3pt

    3. Vector bundles on \cyt s
\vskip3pt

    4. Mapping toric data to vector bundle data
\vskip3pt

\hskip10pt 4.1 {\it The general technique} 

\hskip10pt 4.2 {\it Two examples}

\hskip10pt 4.3 {\it A lower bound for $\eta$}

\hskip10pt 4.4 {\it Three generation models}
\vskip3pt

    5. Fivebranes and extremal transitions
\vskip3pt

    6. Discussion
\vskip 3pt 

    Acknowledgements
\vskip 3pt

    References 
\newpage

\pageno=1
\headline={\ifproofmode\hfil\eightrm draft:\ \today\
\hourandminute\else\hfil\fi}
\footline={\rm\hfil\folio\hfil}
\section{intro}{Introduction}
F-Theory/Heterotic duality~\REFS\Vafa{C.~Vafa, Nucl. Phys. {\bf B469} (1996)
403, hep-th/9602065.}
\REFSCON\rMVI{D.~R.~Morrison and C.~Vafa, Nucl. Phys. {\bf B473} (1996) 74,
hep-th/9602114.}
\REFSCON\rMVII{D.~R.~Morrison and C.~Vafa, Nucl. Phys. {\bf B476} (1996) 437, 
hep-th/9603161.}
\REFSCON\rBer{M.~Bershadsky {\it et al\/.}, Nucl. Phys. {\bf B481} (1996) 215, 
hep-th/9605200.}
\refsend\  provides a useful way of studying nonperturbative
string theory. In its original form, it states that F-theory compactified on an
elliptic $K3$ surface is dual to heterotic string theory on $T^2$ with some
choice of vector bundle $V$, schematically, 
$${{\rm Het}[T^2,V]={\rm F}[K3]}.
\eqlabel{8dim}$$
In particular, the unbroken gauge group can be read off from the singularities
of the elliptic fibration structure of the $K3$. Equation~\eqref{8dim} is a 
statement about eight dimensional theories. We obtain 
lower dimensional versions of this duality by further
compactification, and, using adiabatic arguments,
applying the duality ``fibrewise''. For instance, in six dimensions (by
further compactification of both sides of the above equation on a $\IP^1$),
we arrive at the well known duality relation
$${\rm Het}[K3,V]={\rm F}[{\ca M}_V],
\eqlabel{6dim}$$
where ${\ca M}_V$ is an elliptic Calabi-Yau threefold which depends upon the
choice of vector bundle. A very large class of such \cyt s can be realised 
as hypersurfaces in toric varieties, and one can then establish a dictionary
between the toric data and the heterotic data, including the gauge and matter
content of the corresponding low dimensional
effective field theories, which have $N=1$ supersymmetry in six dimensions
(see, for example,~
\REFS\CF{P.~Candelas and A.~Font, Nucl. Phys. {\bf B511} (1998) 295,
hep-th/9603170.}
\REFSCON\CPRbig{P.~Candelas, E.~Perevalov and G.~Rajesh,
Nucl. Phys. {\bf B507} (1997) 445, hep-th/9704097.}
\REFSCON\CPRmat{P.~Candelas, E.~Perevalov and G.~Rajesh,
Nucl. Phys. {\bf B519} (1998) 225, hep-th/9707049.}
\refsend\ and references therein).

If we were instead to compactify further on a (complex) surface $B_2$,
we obtain the phenomenologically interesting duality between $N=1$ theories in
four dimensions
$${{\rm Het}[Z, V]={\rm F}[X]},
\eqlabel{4dim}$$
with $Z$ an elliptic \cyt\ with base $B_2$ and $X$ an elliptic \cyf\ with a 
three dimensional base $B_3$ which is a $\IP^1$ bundle over $B_2$ (or a blowup
thereof).

To better understand this duality, we first need a general
procedure for constructing vector bundles on elliptic \cyt s, and then we need
to map the fourfold data to the corresponding bundle data. The first of these
questions was addressed in~
\REFS\FMW{R.~Friedman, J.~Morgan and E.~Witten, Comm. Math. Phys. {\bf 187}
(1997) 679, hep-th/9701162.}
~\REFSCON\BJPS{M.~Bershadsky, A.~Johansen, T.~Pantev and V.~Sadov,
Nucl. Phys. {\bf B505} (1997) 165, hep-th/9701165.}
\refsend .
In this paper, we address the second question. Specifically, given
a \cyf\ as a hypersurface in a toric variety, we show to read off the data
necessary to construct the bundle on the heterotic side. We can then count the
number of fivebranes on the heterotic side which wrap the elliptic fibre and
match them to the number of threebranes on the F-theory side which is related
to the Euler number of the fourfold by tadpole anomaly
cancellation~\Ref\SVW{S.~Sethi,
C.~Vafa and E.~Witten, Nucl. Phys. {\bf B480} (1996) 213,
hep-th/9606122.}, providing the first nontrivial test of the map between
toric data and bundle data.
Next, we count the
vector bundle moduli~\cite{\FMW} (these have also been discussed
in~\REFS\CD{G.~Curio and R.~Donagi, Nucl. Phys. {\bf B518} (1998) 603,
hep-th/9801057.}
~\REFSCON\AC{B.~Andreas and G.~Curio, Phys. Lett. {\bf B417}
(1998) 41, hep-th/9706093.}
~\REFSCON\ACL{B.~Andreas, G.~Curio and D.~L\"ust, Nucl. Phys. {\bf B507}
(1997) 175, hep-th/9705174.}
\refsend  ) and match them to the Hodge numbers (specifically,
$h_{31}$) of the \cyf , providing the second nontrivial test of our map.
Using our prescription, we will
show how to construct \cyf s that yield 3 generation models with GUT groups,
following~\Ref\Cur{G.~Curio, Phys. Lett {\bf B435} (1997) 39, 
hep-th/9803224.}. We will then address the question of
heterotic fivebranes wrapping curves in the base. We show that the F-theory
dual of this situation consists of blowing up the corresponding curves in the
fourfold base into ruled surfaces. In particular, we find that when the
fivebranes wrap different curves in the heterotic base which nevertheless lie
in the same homology class, the F-theory duals generally have different
numbers of blowup modes, and hence different Hodge numbers. However, since the
rest of the bundle data are the same, the Euler numbers are unchanged. This
then raises the possibility of following a topology changing extremal
transition on the fourfold in terms of degenerations of curves on the
heterotic side. It is worth emphasizing here that our analysis will be purely
classical. Quantum corrections will not be considered in this work.

The rest of this paper is organized as follows. In \SS{2}, we summarize some
relevant results in toric geometry. In \SS{3}, we briefly discuss the
construction of vector bundles by Friedman, Morgan and Witten~\cite{\FMW}. 
In \SS{4.1}, we describe our procedure for reading off the the bundle data,
specifically, the $\eta$ class from the polyhedra of the fourfold, and give
two examples in \SS{4.2}. In \SS{4.3}, we find a lower bound on the $\eta$
class imposed by the fourfold geometry, which is related to the stability of
the corresponding vector bundle. In \SS{4.4}, we discuss the construction of
three generation models with GUT groups, and
provide an example of such a model. In \SS{5}, we study fivebranes wrapping
curves in the base of the heterotic \cyt , and provide an example
of the topology changing transition mentioned above. \SS{6} concludes with a
discussion of our findings. 

During the preparation of this paper, we became aware of the work of
Donagi, Lukas, Ovrut and Waldram~\Ref\DLOW{R.~Donagi, A.~Lukas,
B.~A.~Ovrut and D.~Waldram, hep-th/9811168.}
which is related to the material
presented in \SS{5}. They analyse the moduli space of fivebranes from an
M-theory perspective, and find that when the cohomology class of the
fivebranes corresponds to a reducible divisor, the moduli space of these
fivebranes has several components. In our paper, we show that the
F-theory duals consist of fourfolds with different Hodge numbers, but the same
Euler number. Thus, the two results seem to complement to each other. After
this work was complete, we received a
preprint~\Ref\BM{P.~Berglund and P.~Mayr, hep-th/9811217.}
which has some overlap with this work.

\newpage  

\section{toric}{Some results in toric geometry}
In this section we briefly summarize some results in toric geometry which will
be relevant to our discussion.
A large class of \cym s can be realised as hypersurfaces in toric varieties,
and are described, using Batyrev's construction
\REFS\rBat{V.~Batyrev, Duke Math. Journ. {\bf 69} (1993) 349.}
\REFSCON\rCOK{P.~Candelas, X.~de la Ossa and S.~Katz,
Nucl. Phys. {\bf B450} (1995) 267,\\ hep-th/9412117.}
\refsend , by a dual pair $(\D, \nabla)$ of reflexive polyhedra. The
polyhedron $\D$ is called the Newton Polyhedron, and describes the monomials
in the equation describing the \cym\ as a hypersurface in the toric variety.
The dual polyhedron $\nabla$ describes the fan of the corresponding toric
variety. The Hodge numbers of the \cym\ are then obtained using the following
formulas. 

For \cyt s, the only independent Hodge numbers are $h_{11}$ and $h_{12}$, which
are given by
$$\eqalign{&h_{21}={\rm pts}(\D) \ -\hskip-10pt
\sum_{{\rm codim}(\theta)=1}\hskip-10pt{\rm int}(\theta) \ +\hskip-10pt
\sum_{{\rm codim}(\theta)=2}\hskip-10pt{\rm int}(\theta)
{\rm int}(\tilde{\theta}) \ - \ 5,\cropen{8pt}
&h_{11}={\rm pts}(\nabla) \ -
\hskip-10pt\sum_{{\rm codim}(\tilde{\theta})=1}\hskip-10pt{\rm int}
(\tilde{\theta}) \ +\hskip-10pt
\sum_{{\rm codim}(\tilde{\theta})=2}\hskip-10pt
{\rm int}(\tilde{\theta}){\rm int}(\theta) \ - \ 5}
\eqlabel{hodgethree}$$
where pts$(\D)$ denotes the number of integral points of $\D$, int$(\theta)$
stands for the number of integral points interior to a face $\theta$ and 
similar quantities pts($\nabla$) and int$(\tilde{\theta})$ are defined for
$\nabla$.
Equation \eqref{hodgethree} expresses the number of deformations
of complex structure and \K\ classes in terms of the number of
points of the polyhedra. The terms in these expressions that involve
codimension-1 faces account, in the case of $h_{21}$, for the freedom to make
redefinitions of the homogeneous variables, and in the case of $h_{11}$, for
the singularities of the toric variety which do not intersect the
hypersurface. The third terms in both
equations are `correction' terms, the numbers of deformations of the
corresponding hypersurface which are not visible torically. (Note that in
many cases it turns out to be possible to add a certain number of
points to the polyhedron under consideration so that the correction vanishes.)

Similarly, for \cyf s, the only independent Hodge numbers are $h_{11},h_{31}$
and $h_{21}$. The fourth nontrivial Hodge number $h_{22}$ is in fact
determined from
$$\chi=48 + 6(h_{11}+h_{31}-h_{21})=4+2(h_{11}+h_{31}-2h_{21})+h_{22},
$$
which also determines the Euler number. The expressions for $h_{11},h_{31}$
and $h_{21}$ are
$$\eqalign{&h_{31}={\rm pts}(\D) \ -\hskip-10pt
\sum_{{\rm codim}(\theta)=1}\hskip-10pt{\rm int}(\theta) \ +\hskip-10pt
\sum_{{\rm codim}(\theta)=2}\hskip-10pt{\rm int}(\theta)
{\rm int}(\tilde{\theta}) \ - \ 6,\cropen{8pt}
&h_{11}={\rm pts}(\nabla) \ -
\hskip-10pt\sum_{{\rm codim}(\tilde{\theta})=1}\hskip-10pt{\rm int}
(\tilde{\theta}) \ +\hskip-10pt
\sum_{{\rm codim}(\tilde{\theta})=2}\hskip-10pt
{\rm int}(\tilde{\theta}){\rm int}(\theta) \ - \ 6,\cropen{8pt}
&h_{21}=\sum_{{\rm codim}(\tilde{\theta})=3}\hskip-10pt
{\rm int}(\tilde{\theta}){\rm int}(\theta)} 
\eqlabel{hodgefour}$$
where pts($\D$), pts($\nabla$), $\theta$, $\tilde{\theta}$, int($\theta$) and
int($\tilde{\theta}$) are defined as before.

In this paper, we will mainly be interested in \cys\ which are elliptic
fibrations. For instance, we will consider heterotic compactifications on
\cyt s that are elliptically fibred over the Hirzebruch surface ${\IF}_m$. 
Then the starting
point is the hypersurface in the toric variety defined by the data displayed
in Table~\tabref{scaling}~\cite{\rMVI}. Namely, start with homogeneous 
coordinates $s,t,u,v,x,y,w$, remove the loci $\{s=t=0\}$, $\{u=v=0\}$,
$\{x=y=w=0\}$, take the quotient by three scalings $(\l,\m,\n)$ with the
exponents shown in Table~\tabref{scaling} and restrict
to the solution set of (homogeneous version of) the Weierstrass
equation~\eqref{Wei}
$$y^2=x^3+f(z,z^{\prime})x+g(z,z^{\prime}),
\eqlabel{Wei}$$
where $z$ and $z^{\prime }$ are affine coordinates on the base.
$$\vbox{\offinterlineskip\halign{
&\strut\vrule height 12pt depth 6pt #&\hfil\quad$#$\quad\vrule
&\hfil \quad$#$\quad&\hfil\qquad$#$\quad&\hfil\qquad$#$\quad
&\hfil\qquad$#$\quad&\hfil\qquad$#$\quad&\hfil\qquad$#$\quad
&\hfil\qquad$#$\quad\vrule&\quad$#$\quad\hfil\vrule\cr
\noalign{\hrule}
&&s&t&u&v&x&y&w&\hfil\hbox{degrees}\cr
\noalign{\hrule\vskip3pt\hrule}
&\l&1&1&\hidewidth{m}&0&\hidewidth{2m{+}4}&\hidewidth{3m{+}6}&0&6m+12\cr
&\m&0&0&1&1&4&6&0&12\cr
&\n&0&0&0&0&2&3&1&6\cr
\noalign{\hrule}
}}
$$
\nobreak\tablecaption{scaling}{The scaling weights of the elliptic
fibration over $\IF_{m}$.}
\bigskip

Similarly, we can construct \cy\ fourfolds that are elliptically fibred over
the generalized Hirzebruch surface ${\IF}_{mnp}$, which is a ${\IP}^1$ bundle
over ${\IF}_m$. The scaling weights are given in Table~\tabref{scalingtwo}.

$$\vbox{\offinterlineskip\halign{
&\strut\vrule height 12pt depth 6pt #&\hfil\quad$#$\quad\vrule
&\hfil \quad$#$&\hfil\quad$#$
&\hfil \quad$#$&\hfil\quad$#$&\hfil\quad$#$
&\hfil \quad$#$\qquad&\hfil\qquad\qquad$#$\qquad&\hfil\qquad\qquad$#$
&\hfil\quad$#$\quad\vrule&\quad$#$\quad\hfil\vrule\cr
\noalign{\hrule}
&&q&r&s&t&u&v&x&y&w&\hfil\hbox{degrees}\cr
\noalign{\hrule\vskip3pt\hrule}
&\k&1&1&\hidewidth{m}&0&\hidewidth{p}&0&\hidewidth{2(m{+}p){+}4}&\hidewidth{3(m{+}p){+}6}&0&6(m+p)+12\cr
&\l&0&0&1&1&\hidewidth{n}&0&\hidewidth{2n{+}4}&\hidewidth{3n{+}6}&0&6n+12\cr
&\m&0&0&0&0&1&1&4&6&0&12\cr
&\n&0&0&0&0&0&0&2&3&1&6\cr
\noalign{\hrule}
}}
$$
\nobreak\tablecaption{scalingtwo}{The scaling weights of the elliptic
fibration over $\IF_{mnp}$.}
\bigskip
For the manifolds described above, the statement of the duality
relation~\eqref{4dim} is that F-theory compactified on a \cyf\ which is an
elliptic fibration over $\IF_{mnp}$ is dual to heterotic string theory
compactified on a \cyt\ which is elliptically fibred over $\IF_m$ with a
vector bundle governed by the data $n$ and $p$ of the fourfold. We shall, in
this paper, attempt to make precise this relation between the vector bundle
and the fourfold.

Toric geometry also encodes in a natural way the fibration structure (if any)
of the \cym s. The authors of~\Ref\rAS{A. Avram, M. Kreuzer, M. Mandelberg
and H. Skarke, Nucl. Phys. {\bf B494} (1997) 567, hep-th/9610154.} state this
for \cys\
that are described by reflexive polyhedra, the integral points of the polyhedra
being points in a lattice $\L$. It has been shown there that in order for
a \cy\ $n$-fold to be
a fibration with generic fiber a \cy\ $(n-k)$-fold it is necessary and 
sufficient that\Footnote{We denote, as is standard, the lattice dual to $\L$
(where $\D$ lives) by $V$, and its real extension by $V_{\IR}$.} 
\item\ {(i) There is a projection operator $\P$: $\L\rightarrow \L_{n-k}$,
where 
$\L_{n-k}$ is an $n-k$ dimensional sublattice, such that $\P(\D)$ is a
reflexive polyhedron in $\L_{n-k}$, or}
\item\ {(ii) There is a lattice plane in $V_{\IR}$ through the origin 
whose intersection with $\nabla$ is an $n-k$ dimensional reflexive polyhedron,
{\it i.e.\/} it is a slice of the polyhedron.}

\noindent (i) and (ii) are equivalent conditions.
If (i) or (ii) hold there is also a way to see the base of the fibration
torically~\REFS\rSp{M.~Kreuzer and H.~Skarke, J. Geom. Phys. {\bf 26} (1998)
272, hep-th/9701175.}\refsend .
The hyperplane $H$ generates an $n-k$ dimensional sublattice of $V$. Denote
this lattice $V_{\rm fiber}$. Then the quotient lattice 
$V_{\rm base}=V/V_{\rm fiber}$ is the lattice in which the fan of the base 
lives. The fan itself can be constructed as follows. Let $\P_B$ be a projection
operator acting in $V$ such that it projects $H$ onto
a point. Then
$\P_B(V)=V_{\rm base}$. When $\P_B$ acts on $\nabla$ the result is a $k$ 
dimensional
set of points in $V_{\rm base}$ which gives us the fan of the base if we draw
rays through each point in the set.

Suppose now that we are given an elliptic \cyt. The theorem of~\cite\rAS\
tells us that in this case it is possible to find a two-dimensional hyperplane 
$H$ in $V_{\IR}$ through the origin such that its intersection with 
$\nabla$ is a two-dimensional reflexive polyhedron representing the typical 
fiber. Let us denote it by $\nabla^{\ca{E}} =\nabla\cap H$. 
Projecting $\nabla$ with $\P_B$ such that $\P_B(\nabla^{\ca{E}})=(0,0)$ yields
a set of 
points living in a two-dimensional lattice which is what we call 
$V_{\rm base}$. Drawing a ray from the origin $(0,0)$ through every 
other point gives us the fan of the base. Note that a ray may pass through
more than one point and hence the number of rays, or one-dimensional cones,
is generically less than the number of non-zero points in 
$V_{\rm base}$. For elliptic \cyf s, the same picture again holds, except that
the base is now a three dimensional toric variety.

In general, the elliptic fibre can degenerate over the divisors in the base. 
The singularity over each divisor in the base gives rise to a factor of
the total gauge group. The method for reading off the singularity structure,
and hence the total gauge group, was proposed in~\cite{\CPRbig}.
For \cyf s, there is a subtlety due to the presence of
a number (generically $\chi\over24$) of threebranes~\cite{\SVW},
required for anomaly
cancellation. If the threebranes were to coincide with any of the sevenbranes
wrapping the singularities, they would behave like instantons and break the
observed gauge group to a smaller group~\cite{\BJPS}\Ref\DM{K.~Dasgupta
and S.~Mukhi, Phys. Lett. {\bf B398} (1997) 285, hep-th/9612188.}.
Generically, however, the threebranes
are located at points of the base where the elliptic fibre is smooth, and
thus do not break the observed group. For the purposes of this paper, we will
assume that the threebranes are indeed generic, and determine the gauge group
from the singularities of the elliptic fibration structure.    
\newpage
\section{bundles}{Vector bundles on \cyt s}
In this section we summarize relevant aspects of the work of
Friedman, Morgan and Witten~\cite{\FMW} on vector bundles. We refer the
interested reader to that work for more details. 

For the purposes of this paper, we will only consider $SU(N)$ bundles and
$E_8$ bundles, although our results should apply to other bundles as well.
Friedman, Morgan and Witten construct $SU(N)$ bundles with $c_1(V)=0$ from
the spectral cover
$C$, which is described as follows. Consider a
semistable $SU(N)$
bundle on an elliptic curve $E$, which has a distinguished point $p$,
the ``origin''. This determines a vector bundle $V$ which 
splits into a sum of $N$ line bundles, $V=\oplus_{i=1}^{N}{\ca N}_i$.
The fact that the bundle is $SU(N)$ means that the product of the ${\ca N}_i$
is the trivial line bundle, and the fact that it is semistable implies that
the ${\ca N}_i$ are all of degree zero. 

For any degree zero line bundle ${\ca N}_i$, there is a unique point $Q_i$,
such that ${\ca N}_i$ has a holomorphic section which vanishes only
at $Q_i$ and has a pole only at $p$. Thus $V$ is determined by the $N$
points $Q_i$ on $E$. Since the product of the ${\ca N}_i$ is trivial, the sum
(using addition with respect to the group law on $E$) of the $Q_i$ is zero.
Conversely, for any point $Q_i$ in $E$, there is a unique line bundle
${\ca N}_i = {\ca O}(Q_i)\otimes{\ca O}(p)^{-1}$, so every $N$-tuple of
points in $E$ (which add up to zero) determines a semistable $SU(N)$ bundle.

Now, for an elliptic \cyt\ , we can ``fibre'' the above bundle construction
over the base $B_2$ of the elliptic fibration, obtaining an $N$-fold cover of
the
base. This is the spectral surface $C$ of the bundle. The spectral surface is 
actually a section of ${\ca O}(\sigma)^N\otimes{\ca M}$, where $\sigma$ is the
zero section of the elliptic fibration (corresponding to a global choice of
reference point $p$), and ${\ca M}$ is an arbitrary line bundle over $B_2$,
with $c_1({\ca M})=\eta$. The class $\eta$ is the single most important
ingredient in the construction of the vector bundle.

Reconstructing the bundle from the spectral cover involves the
Poincar\'e line bundle.
We will not go into this topic here, but refer the
reader to Ref.~\cite{\FMW}. We simply note here a result of~\cite{\FMW} that
in general one must twist by a line bundle ${\ca N}$ over $C$ in order to
reconstruct a specific $SU(N)$ bundle\Footnote{A further generalization of this
is mentioned in~\cite{\Cur}, but we will not consider this here.}.
When $H^{1,0}(C)=0$, the classification of such line bundles on $C$ is
discrete, and ${\ca N}$ is uniquely determined by its first Chern class. In
more general situations, we will also need to specify an element
of the intermediate Jacobian $H^3(X,{\bf R})/H^3(X,{\bf Z})$, where $X$ is the 
dual \cyf . However, when $h^3(X)$ is zero, which is the case for many
\cyf s, this complication does not
arise~\cite{\FMW ,\CD}. In fact, all the examples studied in this paper satisfy
$h^3(X)=0$, although our methods will also be applicable to the more general
cases.

The most general form of the line bundle ${\ca N}$ is~\cite{\FMW}
$${\ca N}=K_C^{1/2}\otimes K_B^{-1/2}\otimes ({\ca O}(\sigma)^N\otimes
{\ca M}^{-1}\otimes {\ca L}^N)^{\l},
\eqlabel{Nbundle}$$
with $c_1({\ca L})=c_1(B_2)$ (for elliptic \cyt s) and
suitable $\l$. If the square root $K_C^{1/2}\otimes K_B^{-1/2}$
does not exist, then one cannot set $\l$ to zero, and must in fact choose
$\l$ half-integral. In fact the only circumstance in which
$K_C\otimes K_B^{-1}$ has a square root is if
$$\eqalign{&N \equiv 0 \; {\rm mod} \; 2\cr
&\eta \equiv c_1({\ca L}) \; {\rm mod} \; 2.\cr}
\eqlabel{zeroN}$$ 
Equation~\eqref{Nbundle} implies
$$c_1({\ca N})={1\over2}(N\sigma + \eta + c_1(B_2)) + \g,
\eqlabel{chernN}$$
with
$$\g=\l(N\sigma - \eta + Nc_1(B_2)).
\eqlabel{gamma}$$
Furthermore, $\t$-invariant bundles (where $\t$ is the involution on
the elliptic fibre) have $\g=0$.

Now, the Chern classes of the corresponding $SU(N)$ bundle $V$ can be computed
to be
$$\eqalign{&c_2(V)= \eta\sigma - {c_1({\ca L})^2(N^3-N)\over24}
- {N\eta(\eta - Nc_1({\ca L}))\over 8} - {\p_\ast(\g^2)\over2},\cr
&c_3(V)=2\l\eta(\eta-Nc_1(B_2)),\cr}
\eqlabel{chernV}$$
with
$$\p_\ast(\g^2)=-\l^2N\eta(\eta-Nc_1({\ca L})).
\eqlabel{gammasq}$$ 
The second of Equations~\eqref{chernV} was worked out in~\cite{\Cur}. Note
that ${1\over2}c_3(V)$ is the net generation number, so we see that the only
way to obtain chiral matter is to have non-$\t$-invariant bundles.  

We do not have a spectral cover description of $E_8$ bundles. Semistable
$E_8$ bundles are constructed in Ref.~\cite{\FMW} by the
method of parabolics.
We will not explore the details of this construction here, but merely note that
this method only yields $\t$-invariant bundles.
Since $E_8$ bundles are real, the third Chern class is trivial. The second
Chern class is given by
$$c_2(V)=60(\eta\sigma - 15\eta^2 + 135\eta c_1({\ca L}) - 310 c_1({\ca L})^2).
\eqlabel{chernE}$$

Before proceeding to the next section, we pause to note some important
constraints on the $N=1$ F-theory/Heterotic vacua in four dimensions.
It was shown in~\cite{\SVW} that tadpole
anomaly cancellation requires that the F-Theory vacuum include $\chi(X)/24$
threebranes whose worldvolume is the uncompactified spacetime, and this
requires the heterotic dual to have an equal number of fivebranes wrapping
the elliptic fibre~\cite{\AC}.
This statement is modified in the presence of the flux of the four form field
strength $G$ of the three form gauge field of eleven dimensional
supergravity. Also, the location of
some of the threebranes may coincide with those of the sevenbranes wrapping
divisors in the base $B_3$ over which the elliptic fibre degenerates. These
threebranes then behave like instantons, breaking the observed gauge group to
a smaller group. For the purposes of this paper, we will assume that the
locations of the threebranes are sufficiently generic, so that the
singularities of the fibration do in fact yield the true gauge group. Then
the tadpole anomaly
cancellation condition is
$${\chi\over24}= n_3 + \int_{X}{G^2\over2}.
\eqlabel{tadpole}$$
It was also argued in~\Ref\Wit{E.~Witten, J. Geom. Phys. {\bf 22}
(1997) 1, hep-th/9609122.} that $G$ is quantized in half integer
units. This suggests a natural relation between the four flux and the $\g$
class, which has a similar quantization, and it was argued in
Ref.~\cite{\CD} that in fact $\int_{X}{G^2\over2}=-{\p_\ast(\g^2)\over2}$. 

The other constraint is the general heterotic anomaly cancellation condition
$$\l(V_1)+\l(V_2)+[W]=c_2(TZ),
\eqlabel{hetanomaly}$$
where $\l(V)$ is the fundamental characteristic class of the vector bundle $V$
(which is $c_2(V)$ for $SU(N)$ bundles and $c_2(V)/60$ for $E_8$ bundles),
$[W]$ is the cohomology class of the fivebranes, and $TZ$ is the tangent
bundle of $Z$.
Furthermore, for the models that we consider,
$c_2(TZ)= 12c_1(B_2)\sigma + 11c_1^2(B_2) + c_2(B_2)$. Thus, we can integrate
Equation~\eqref{hetanomaly} over the base $B_2$ of the heterotic threefold, and
arrive at the number of fivebranes wrapping the elliptic fibre. Thus, we
arrive at a non-trivial consistency check for any map relating \cyf s and
vector bundles
--- for any \cyf , the corresponding vector bundle will be such as to yield
a number
of fivebranes wrapping the elliptic fibre by Equation~\eqref{hetanomaly},
which must equal the number of threebranes in Equation~\eqref{tadpole}.
The map that we propose in the next section yields models that do in fact
satisfy this constraint.

We can also relate the bundle moduli to the Hodge numbers of the fourfold.
The bundle moduli consist of even (\ie\ $\t$-invariant) and odd chiral
superfields, of which there are $n_e$ and $n_o$, respectively. So far, there is
no known method of computing $n_e$ and $n_o$, but an index theorem
in~\cite{\FMW} allows us to compute the difference
$I= n_e - n_o$. From~\cite{\CD ,\AC}, we can relate these to the Hodge numbers
$h_{21}$ and $h_{31}$ of the fourfold as follows
$$\eqalign{&h_{21}=n_o,\cr
&h_{31}= h_{21}(Z) + n_e + 1 = h_{21}(Z) + I + h_{21} + 1,\cr}
\eqlabel{bundlehodge}$$
where unspecified Hodge numbers refer to the fourfold $X$.
The index $I$ is given by~\cite{\FMW}
$$I=-{1\over2}\sum_{i=0}^{3}(-1)^i{\rm Tr}_{H^i(Z,Ad(V))}\t.
\eqlabel{Idef}$$
For $\t$-invariant bundles, we get
$$I=r-4\int_{\sigma}{\l(V)|_{\sigma}}-3\int_Z{c_1(\ca L)\l(V)},
\eqlabel{Itau}$$
where $r$ is the rank of the structure group of the bundle and
$\sigma |_\sigma = -c_1(\ca L)|_\sigma$. This formula
cannot be applied for bundles that are not $\t$-invariant. For
non-$\t$-invariant $SU(N)$ bundles, we compute $I$ using another
formula of~\cite{\FMW}
$$I=-1 + \int_B{e^{\eta}(1+e^{-2c_1(\ca L)}+e^{-3c_1(\ca L)}+\dots+
e^{-Nc_1(\ca L)}){\rm Td}(B)},
\eqlabel{ISUN}$$
where Td is the Todd class, defined for any complex manifold $W$ by
$${\rm Td}(W)=1+{c_1(W)\over2}+{c_2(W)+c_1^2(W)\over12}+\dots
\eqlabel{Todd}$$
In fact, it was shown in~\cite{\FMW} that Equation~\eqref{ISUN} agrees with
Equation~\eqref{Itau} for $\t$-invariant bundles. 
Using these formulas, we obtain our second consistency check of the map
that we propose in the next section, and all the models that we study satisfy
this constraint.

We have not yet discussed the third independent Hodge number of the fourfold,
namely, $h_{11}$. A formula in~\cite{\AC} gives
$$h_{11}=h_{11}(Z)+1+{\rm rank}(G),
\eqlabel{h11old}$$
where rank($G$) is the rank of the unbroken non-abelian gauge group. However,
when we have fivebranes wrapping curves in the heterotic base $B_2$ as in
\SS{5}, this formula will have to be modified to include the number of blowups
of the base $B_3$. The correct formula (when $Z$ has a smooth Weierstrass
fibration, which will be true of all the models we study in this paper) by
analogy with the six dimensional situation, is
$$h_{11}(X)=1+h_{11}(B_3)+\hbox{rank}(G),
\eqlabel{h11new}$$
where $G$ is the unbroken gauge group.

\newpage
\section{map}{Mapping toric data to vector bundle data}
\subsection{The general technique}
Recall that we can consider the \cyf\ to be a $K3$ fibration over $B_2$. The
$K3$ fibre itself has an elliptic fibration compatible with the elliptic
fibration structure of the \cyf . In general, the elliptic fibre can
degenerate over several divisors in the threefold base of the \cyf , leading
to enhanced gauge symmetry. For the purposes of this paper, we will only
consider the situation when these singularities lie in the $K3$ fibre. This
is the analogue in four dimensions of the six dimensional case when the gauge
group was purely perturbative (from a heterotic perspective), \ie\ a subgroup
of the heterotic $E_8\times E_8$ gauge group. The unbroken gauge symmetry was
then the commutant of the structure group of the vector bundle, and was
related to the singularity type by identifying a singularity of type
{\ss ADE} with the corresponding {\bf ADE} group. Using the
adiabatic argument, therefore, we conclude that in the fourfold situation (if
all the singularities of the elliptic fibration lie in the $K3$ fibre), the
gauge group that we read off from the singularities is just the commutant in
$E_8\times E_8$ of the structure group of the vector bundle. The assumption
that all the singularities of the elliptic fibration lie in the $K3$ fibre
means in particular that the \cyt\ on the heterotic side has a {\sl smooth\/}
Weierstrass fibration.

We still need to specify the bundles themselves. For this, we need to specify,
among other things,
the $\eta$ and $\g$ classes. We relate them to the toric data as follows. The
base of the $K3$ fibre is a $\IP^1$ which is precisely the $\IP^1$ fibre of
$B_3$ over $B_2$. From the discussion in \SS{2} (since the $K3$ fibration of
the \cyf\ is compatible with its elliptic fibration), the fan of this
$\IP^1$ is seen as a slice through the origin in the fan of $B_3$. Now
the fan of $\IP^1$ consists simply of two rays, $R_1$ and $R_2$, opposite each
other. These correspond to divisors in the base 
$B_3$ (see Table~\tabref{fmnp}), and the singularities
${\ss G_1}$ and ${\ss G_2}$ of the elliptic
fibration over $R_1$ and $R_2$ (which are read off from the preimages of $R_1$
and $R_2$ under the map which projects the polyhedron $\nabla$ of the \cyf\
onto
the fan of the base) give rise to the gauge group $G_1\times G_2$ which is
the commutant in $E_8\times E_8$ of the structure group $V_1\times V_2$ of the
vector bundle. Thus $V_1$ and $V_2$ are naturally associated to the rays $R_1$
and $R_2$ in the fan of $B_3$. 

For each $R_i$, we define a divisor $t_i$ in $B_2$ as follows. Due to the
linear relations of the fan~\Ref\Ful{W. Fulton, Introduction to Toric
Varieties, Princeton University Press, 1993.}, we have
$R_2 = R_1 + \sum A_jD_j$ where the sum
runs over all the other divisors $D_j$ in $B_3$ and the $A_j$ are integers.
Now, $R_1 . R_2 = 0$, so $R_1 . (R_1 + \sum A_jD_j) = 0$. 
Clearly, in this expression, we can restrict the sum to the divisors $D_j$
that actually intersect $R_1$. If $\p(D_j)$ is the image of $D_j$ under the
projection $\p$ from $B_3$ to $B_2$, then we write
$$t_1 =  \sum_{D_j . R_1 \neq 0}{A_j \p(D_j)}.
\eqlabel{tdef}$$
(For experts in toric geometry, $t_1$ is a linear combination of divisors in 
Star($R_1$). A similar expression, this time as a linear combination of
divisors in Star($R_2$), then holds for $t_2$.)

We now define $\eta(V_i)$ for the vector bundles $V_i$ as
$$\eta(V_i) = 6c_1(B_2) - t_i, (i=1,2).
\eqlabel{etas}$$
We claim that this definition of $\eta$ gives us precisely the $\eta$ of
Friedman, Morgan and Witten~\cite{\FMW} where $\eta$ was defined
in terms of a class $t$ that satisfied $r(r+t)=0$ for the class $r$ of the
zero section of $B_3$ over $B_2$. In our construction, we identify $R_1$ with
$r$, and then our definition of $t_1$ matches that of $t$. When $B_3$ is a 
$\IP^1$ bundle over $B_2$ (and not a blowup thereof), then $t_2$ is simply
$-t_1$, and our definitions reproduce the definitions in~\cite{\FMW}.
However, our definitions generalize naturally to the case when
$B_3$ is a blowup of a $\IP^1$ bundle over $B_2$ which will become important in
\SS{5}.

Since the definitions of $t$ and $\eta$ above are rather abstract,
we illustrate
them with the following example. Consider the situation when the heterotic
\cyt\ is elliptically fibred over $\IF_m$, while the dual \cyf\ is fibred over
$\IF_{mnp}$. The fan of $\IF_{mnp}$ is generated by rays through the
points in Table~\tabref{fmnp}, where we have also labeled the corresponding
divisors.
\topinsert
$$\vbox{\offinterlineskip\halign{
&\strut\vrule height 12pt depth 6pt #&\hfil\quad$#$\quad\hfil\vrule
&\hfil \quad$#$\quad\hfil\vrule\cr
\noalign{\hrule}
&\hbox{Divisors}&\hbox{Points}\cr
\noalign{\hrule\vskip3pt\hrule}
&R_1&(0,0,1)\cr
&D_1&(1,m,p)\cr
&D_2&(0,1,n)\cr
&D_3&(0,-1,0)\cr
&D_4&(-1,0,0)\cr
&R_2&(0,0,-1)\cr
\noalign{\hrule}
}}
$$
\nobreak\tablecaption{fmnp}{The points generating the fan of $\IF_{mnp}$.}
\endinsert
\bigskip

Note that the rays $R_1$ and $R_2$ form a slice $(0,0,z)$ of the fan through
the origin, and give the $\IP^1$ fibre of $\IF_{mnp}$, while projecting out
the $R_i$, \ie\ mapping $(x,y,z)$ to $(x,y)$ gives the fan of the
base $\IF_m$. 
The linear relations of the fan~\cite{\Ful} imply that
$$\eqalign{&D_1=D_4\cr
  &D_3=D_2 + mD_1\cr
  &R_2=R_1 + pD_1 + nD_2\cr}
\eqlabel{fmnprels}$$
The first two of these imply that $D_1=D_4=f$, the fibre class, $D_2=C_0$,
the zero section and $D_3=C_\infty$, the infinity section of $\IF_m$ regarded
as a $\IP^1$ bundle over $\IP^1$ (strictly speaking, we should take $\p(D_i)$,
where $\p$ is the projection from $\IF_{mnp}$ to $\IF_m$). Furthermore, the
rays $R_1$ and $R_2$ generate
the fan of the $\IP^1$ fibre of $\IF_{mnp}$ over the base $\IF_m$. Following
the prescription given above, we read off $t_1=pf+nC_0=-t_2$, and thus obtain 
$\eta_1$ and $\eta_2$, which agrees with the results of~\cite{\BJPS}.

Now, for $SU(N)$ bundles, $\g$, the analogue of the four flux, is not
determined by the polyhedron,
since we have to specify the four flux in addition to specifying the
fourfold. However, because of tadpole anomaly cancellation~\eqref{tadpole},
and because the number
of fivebranes wrapping the elliptic fiber on the heterotic side is
non-negative, we find that the $\g$ class cannot be arbitrary, but is often
restricted to a small set of possibilities. Thus, ${1\over 2} c_3(V)$, the net
number of generations, which by a formula of Curio~\cite{\Cur} is related to
the $\g$ class, is also restricted to a small number of possibilities for any
given $\eta$. Thus, we find that we must tune $\eta$ to very special values if
we wish to obtain, say, a model with 3 generations. Note that the $\g$ class is
well defined only for bundles which can be described by the spectral cover
method, including $SU(N)$ bundles. For $E_8$ bundles, the $\g$ class does not
exist\Footnote{I am grateful to E.~Witten for explaining this point.}.
We now give a couple of illustrative examples.
\subsection{Two examples} 
$\underline{\it 4.2.1}\quad$
Consider, for example, the fourfold which is the elliptic fibration over
$\IF_{mnp}$ with $m=1, n=12$ and $p=18$. This has an $E_8$ singularity over
the zero section (\ie\ over the divisor which we have called $R_1$). The
Hodge numbers are\Footnote{The Hodge numbers of all the manifolds discussed in
this paper have been computed using the program {\tt POLYHEDRON}, written by
P.~Candelas.}
$h_{11}=12, h_{31}=27548, h_{21}=0, h_{22}=110284$ and Euler
number $\chi=165408$. Note that $h_{11}=4+\hbox{rank}(E_8)$, which is the
analog of a similar relation in threefolds, which agrees with
Equation~\eqref{h11new}.
Here, and in what follows, we will always use 
the notation of Table~\tabref{fmnp} to describe divisors in the base.

The heterotic \cyt\ is an elliptic fibration over $\IF_1$. This has a 
smooth Weierstrass model, and Hodge numbers $h_{11}=3, h_{21}=243$.
From the fourfold, we see that the structure group
$V_1$ is trivial, while $V_2=E_8$. In this case, since we do not have any $SU$
bundles, we have no bundle analogue for the four flux, hence, we set the total
four flux on the F-theory side to zero.
Clearly, the only possible value for
$\eta(V_1)$ is 0. (A trivial bundle with non-trivial $\eta$ is absurd).
Therefore, we must have $t_1 = -6c_1(\IF_1)$. This is actually the case --- if
$C_0$ and $f$ denote the zero section and fibre class of $\IF_1$, we have
$c_1(\IF_1) = 2C_0 + 3f$, while $t_1$ from the fan of $\IF_{mnp}$ is easily
seen to be $- 12C_0 - 18f$. Also, in our example, $\eta(V_2)=12c_1(\IF_1)$.
Given this data, and knowing that $C_0^2=-1, f^2=0, C_0.f=1$, we can readily 
compute $\int_{B_2}\l(V)$ to be $-6800$. The number of fivebranes, $n_5$, is
$$n_5 = \int_{B_2}c_2(TZ) - \int_{B_2}\l(V)= 92 + 6800 = 6892$$ which is
precisely ${\chi\over 24}$, as would be expected from $n_5=n_3$ and
${1\over 2} G^2=0$. Note that the net number of generations is zero, since we
have a $\tau$-invariant bundle.

The bundle moduli can be computed using Equations~\eqref{chernE}
and~\eqref{Itau}. With $\eta=12c_1(\IF_1)$, we find, after some trivial
algebra, that $I=27304$. Also we have $h_{21}(Z)=243$, so that we obtain
$1+h_{21}(Z)+I+h_{21}(X)=27548=h_{31}(X)$, which satisfies the second
consistency check of our map.

In studying this model, we have obtained an important condition. This is that
if we have a divisor corresponding to unbroken $E_8$, the corresponding
$t$ {\sl must\/} be equal to $-6c_1(B_2)$, so that
$\eta$ vanishes. This is the fourfold analogue of the situation for threefolds
where the self-intersection of any divisor corresponding to unbroken $E_8$ had
to be $-12$, and vice versa. Similarly, we conclude that if $t=-6c_1(B_2)$,
then the elliptic fibration {\sl must \/} have an $\ss E_8$
singularity along the
corresponding divisor, leading to an unbroken $E_8$ gauge group.
This is because the corresponding bundle has trivial $\eta$, and hence must
be trivial. We will have more to say about this shortly.
 
\noindent$\underline{\it 4.2.2}\quad$
Now enforce an $E_6$ singularity along the ray $R_2$, which lies
opposite the
ray $R_1$, by adding points to $\nabla$, as in Ref~\cite{\CF}. The fourfold
has Hodge numbers $h_{11}=18,
h_{31}=1670, h_{21}=0, h_{22}=6796$ and Euler number $\chi=10176$. Note that
$h_{11}$ agrees with Equation~\eqref{h11new}.

On the heterotic side, we again have an elliptic fibration over $\IF_1$, but
we now have an $SU(3)$ bundle with $\eta=12c_1(\IF_1)$ and we are forced to
have a non-trivial $\g$ class from Equation~\eqref{zeroN}. We then have
$\int_{\IF_1}c_2(V) = -332 + \l^2 (1296)$, with $\l = k + {1\over 2}$
for some integer $k$. This gives $n_5 = 424 - 1296\l^2$, and requiring that
$n_5$ be non-negative yields $\l=\pm {1\over 2}$, giving $n_5=n_3=100$ and
${1\over 2}G^2 = 324$. Finally, ${\chi\over 24} = 424 = n_3 + {1\over 2} G^2$, 
as expected. The net number of generations is then ${1\over 2} c_3(V)=\pm 432$,
corresponding to $\l=\pm {1\over 2}$. Note that in this case, the net number
of generations is restricted to two possible values, corresponding to the two
possible values of the $\g$ class. This illustrates the statements made above,
that the $\g$ class, though not specified by the choice of fourfold, is
nevertheless restricted to a small set of values by the condition that
the number of fivebranes be positive.

We can now compute the bundle moduli using Equations~\eqref{chernV}
and~\eqref{ISUN}. After some trivial algebra, we find $I=1426$, and so
$1+h_{21}(Z)+I+h_{21}(X) = 1+243+1426+0 = 1670 = h_{31}(X)$, as expected.
Note that because our $SU(3)$ bundle is not
$\t$-invariant, we cannot use Equation~\eqref{Itau} to compute $I$. In
fact, Equation~\eqref{Itau} gives $I=130$, which violates our consistency
check. This, of course, is not a problem, since Equation~\eqref{Itau} was 
derived for $\t$-invariant bundles and is not expected to hold for
bundles which are not $\t$-invariant.
\subsection{A lower bound for $\eta$}
While studying the first example above, we obtained the condition that
$\eta=0$ must correspond to an $E_8$ singularity. We also mentioned that this
is the analogue of the situation in six dimensions, where an $E_8$ singularity 
implies that the corresponding divisor in the base has
self-intersection $-12$. Now, in six dimensions, for any
gauge bundle $H$ on the heterotic side, we also had to have a minimum number
of instantons (\eg\ we require a minimum of 4 instantons for a $SU(2)$ bundle,
and 10 instantons for an $E_8$ bundle). Since $\eta$ is the four dimensional
analogue of the instanton number, it is natural to wonder if there is a
``minimum'' $\eta$ for any gauge bundle. Since $\eta$ is a divisor, we need to 
define the notion of ``minimum''. We define the ``minimum value''
$\eta_{\hbox{min}}(H)$ for any gauge bundle with structure group $H$
to be such that for any $\eta$
with $\eta_{\hbox{min}}(H) - \eta$ an effective divisor, the singularity
corresponding to (\ie\ enforced by) $\eta$ is worse than $G$, where the
corresponding group $G$ is the commutant of $H$ in $E_8$. 
In particular, if $G$ is a subgroup of $E_8$, then we see that
$\eta_{\hbox{min}}(H)$ must itself be an effective divisor, since $\eta=0$
enforces an $E_8$ singularity, which is worse than $G$.
So we have reason to believe that the notion of $\eta$ is well defined.

Another motivation for the existence of a lower bound on $\eta$ is
provided by the toric data. When we compactify F-theory on an elliptic \cyt\
with base $\IF_m$, reflexivity of the
polyhedron $\nabla$ forces us to add points to $\P^{-1}(C_0)$ (where $\P$ is
the projection from $\nabla$ to the base $\IF_m$), and the greater the value
of $m$, the more points we need to add. These points signal a degeneration
of the elliptic fibre over $C_0$. The greater the value
of $m$, the worse the singularity. Now,
in the case of fourfolds, the class $t$ plays much the same role with respect
to
$R_1$ as $m$ does for $C_0$. Thus, the ``bigger'' the class $t$, the worse the
singularity forced over $R_1$, and the smaller the value of $\eta$. So, for
any given $\eta$, there is a ``minimum'' singularity forced over $R_1$, and
hence a ``minimum'' gauge group. In other words, to any given vector bundle,
there must correspond a minimum value of $\eta$. 

We will  propose an expression for
the lower bound for $\eta$ by analogy with the six
dimensional situation. The argument given below was first put forward
in \SS{6.4} of Ref.~\Ref\ASP{P.~S.~Aspinwall, hep-th/9611137.}.
Consider F-theory compactified on a \cyt\ which is an 
elliptic fibration over $\IF_{m}$. Consider the zero section $C_0$ of
$\IF_{m}$, and let the elliptic fibre degenerate over $C_0$, giving a $G$
singularity. Then the discriminant locus $\D=12c_1(\IF_{m})$ vanishes to order
$\delta(G)$ over $C_0$. The values of $\delta(G)$ for any gauge singularity
are obtained from Tate's algorithm (Ref.~\Ref\Tate{J.~Tate,
in {\it Modular Functions of One Variable IV\/}, Lecture Notes in Math.
vol. 476, Springer-Verlag, Berlin, (1975).}).
If we subtract this contribution from the discriminant,
we expect the remainder $\D^{\prime}=\D - \delta(G)C_0$ to have only transverse
intersections with $C_0$, so that $\D^{\prime}.C_0\ge0$ (otherwise the
singularity over $C_0$ would be worse than $G$). Thus we obtain, since
$c_1(\IF_{m})=2C_0+(m+2)f$, that $m\le{24\over12-\delta(G)}$. Now, the number
of instantons in the corresponding bundle is $12-m$, so we see that the minimum
number of instantons to enforce a $G$ singularity is
$12 - {24\over12-\delta(G)}$. 

For example, the number of instantons that will cause the degeneration
of the fibre to be no worse than $I_1$ (\ie\ no singularity) is 
$\ge 12 - 2{2\over5}$ (since $\delta(I_1)=2$), so we need at least 10
instantons for a smooth fibre, \ie\ we need at least 10 instantons
for an $E_8$ bundle. Similarly, 
the number of instantons required to enforce an $SU(2)$
singularity is $\ge 12 - 2{2\over3}$ (since $\delta(SU(2))=3$), so we must have
at least 10 instantons for an $SU(2)$ singularity (with 9 instantons, we get
$SU(3)$). Turning this around, we can then say
that we must have at least 10 instantons to fill out an $E_7$ bundle.
Similarly, since $\delta(E_7)=9$, the minimum number of instantons for an $E_7$
singularity is $12-8=4$, \ie\ we must have at least 4 instantons for an $SU(2)$
bundle.

We could attempt to derive a lower bound for $\eta$ in the fourfold
case in the same way. For instance, we could analyse the degeneration of the
fibre over $R_1$. We can also approach the problem differently. We
note that the quantity $\eta=6c_1(B_2)-t$, where $B_2$ is the heterotic base,
is analogous to the instanton number
$12-m$~\cite{\FMW}. Basically, we replace the instanton number $k$ by the class
${k\over2}c_1(B_2)$.
So we propose the following ansatz: the minimum $\eta$
for any singularity $G$ is
$$\eta_{\hbox{min}}(H)= (6 - {12\over12-\delta(G)})c_1(B_2),
\eqlabel{etamin}$$ 
where $H$ is again the commutant of $G$ in $E_8$.
For example, if $G=E_8$, since $\delta(E_8)=10$, we find that
$\eta_{\hbox{min}}(E_8)=0$, which is consistent with our previous results.

It is extremely interesting to explore the meaning of this lower bound
for $\eta$. It turns out that it is related to the stability of the 
corresponding vector bundle. The following argument was explained to
me by R.~Friedman. From Corollary 6.3 of~\Ref\FMWII{R.~Friedman, J.~Morgan
and E.~Witten, alg-geom/9709029.}
it follows that at least one of the line
bundles
${\ca M}, {\ca M}\otimes{\ca L}^{-2},\dots,{\ca M}\otimes{\ca L}^{-n}$ must
have a section,
and since  ${\ca L^4}$ or ${\ca L^6}$ has a section, this suggests
$\eta=c_1({\ca M})$ must have a ``lower bound''. (Recall that
$c_1({\ca L})=c_1(B_2)$). 
Furthermore, if ${\ca M}$ is
trivial, then {\SS 5.6} of Ref.~\cite{\FMWII} states that the vector bundle is
usually never stable, so that ${\ca M}$ is usually at least as effective as
${\ca L}^2$. 

In fact, we see that Equation~\eqref{etamin} agrees with the above
statement. For an $SU(2)$ bundle, corresponding to an $E_7$ singularity, we see
that $\eta_{\hbox{min}}(SU(2))=2c_1(B_2)$, corresponding to a line bundle
${\ca M}_{\hbox{min}}={\ca L}^2$. Since $SU(2)$ is essentially the
``smallest'' vector bundle we can have, this means that in general,
${\ca M}$ is at least as effective as ${\ca L}^2$, as stated above.
But Equation~\eqref{etamin} was obtained from purely 
geometric considerations on the \cyf . Thus, we see one of the miracles
of string duality - the elliptic \cyf\ ``knows'' about the stability of
vector bundles on the \cyt !

Before we leave this topic, we mention in passing that $\eta$ is also bounded
from above. For supersymmetric vacua, we need the class of fivebranes to be
effective (otherwise we would have antibranes~\cite{\DLOW}).
Then it follows trivially that
$\sum_{i}{\eta(V_i)} \le12c_1(B_2)$, where we sum over all the bundle factors.

\subsection{Three generation models}
Given the relation between bundle data and fourfold data, we are in a position
to attempt to construct models with three generations and GUT groups. Consider,
for instance, a three generation model with $E_6$ gauge symmetry. This
corresponds to an $SU(3)$ bundle with ${1\over2}c_3(V)=3$. This puts
constraints on the bundle data $\eta$ and $\g=\l\eta(\eta-3c_1(B_2))$, and
one can attempt to find solutions to these constraints.

Some of these solutions are given in~\cite{\Cur}. For $SU(3)$ bundles on
\cyt s with base $\IF_m$, the solution given there is $\eta=f$,
$\l=-{1\over2}$. (Note that $m\le2$ for the heterotic \cyt\ to have a smooth
Weierstrass model.) This solution, is however, impossible, by the lower bound
of the previous section. Since $\delta(E_6)=8$, we have
$\eta_{\hbox{min}}(SU(3))=3c_1(B_2)=6C_0+3(m+2)f>f $, so this solution violates
the lower bound. In fact, this solution also violates the weaker condition that
${\ca M}$ should be at least as effective as ${\ca L}^2$.
Indeed, it is impossible to construct a reflexive
polyhedron with this value of $\eta$ and an $E_6$ singularity ---
the singularity forced over $R_1$ is worse than $E_6$. 

While it is possible to attempt to work out all the solutions to the above
constraints, we will present just one. The elliptic fibration over
$\IF_{1,6,8}$ with an $E_6$ singularity imposed over $R_1$
and with four flux fixed by setting $\l = {1\over2}$
is dual to a heterotic compactification on the smooth \cyt\ over $\IF_1$,
with an $SU(3)$ bundle with $\eta=6C_0+10f$ and $\g$ determined
by $\l = {1\over2}$, as well as a $\tau$-invariant $E_8$ bundle with
$\eta=18C_0+26f$.
The polyhedron $\nabla$ consists of the points shown below:
$$\eqalign{
&{\hbox{\tt(-1,0,0,2,3),(0,-1,0,2,3),(0,0,0,-1,0),(0,0,0,0,-1),(0,0,0,0,0),}}\cr
&{\hbox{\tt(0,0,1,0,0),(0,0,0,0,1),(0,0,1,0,1),(0,0,0,1,1),(0,0,1,1,1),}}\cr
&{\hbox{\tt(0,0,2,1,1),(0,0,0,1,2),(0,0,1,1,2),(0,0,2,1,2),(0,0,-1,2,3),}}\cr
&{\hbox{\tt(0,0,0,2,3),(0,0,1,2,3),(0,0,2,2,3),(0,0,3,2,3),(0,1,6,2,3),(1,1,8,2,3)}}\cr}$$
The \cyf\ has Hodge numbers $h_{11}=10,h_{31}=9231,h_{21}=0,h_{22}=37008$ and
Euler number $\chi=55494$. Note that $h_{11}=4+{\rm rank}(E_6)$. 
Note also that since $h_{21}=0$, the spectral curve of
the $SU(3)$ bundle has $h_{10}=0$, so that the bundle is completely specified
by the data given above. In particular, we find $n_5=n_3=2310$,
${1\over2}G^2=-{\p_{\ast}(\g^2)\over2}=9/4$, and $\chi=24(n_3+{1\over2}G^2)$,
as expected. Finally, this is a 3 generation model: ${1\over2}c_3(SU(3))=3$.
(For computing the net generation number, we ignore the contribution from the
hidden sector bundle, which in our example gives zero anyway.)

Next, let us count the moduli of our bundles.
Using Equations~\eqref{chernE} and~\eqref{Itau}, we find for the $\t$-invariant
$E_8$ bundle, $I_{E_8}=8918$. Also from Equation~\eqref{ISUN}, we find, for 
the non-$\t$-invariant $SU(3)$ bundle,  $I_{SU(3)}=69$, so that
$1+h_{21}(Z)+I_{E_8}+I_{SU(3)}+h_{21}(X) =1+243+8918+69+0 =9231 = h_{31}(X)$,
consistent with our expectations. As in example {\it 4.2.2\/},
note that we cannot use
Equation~\eqref{Itau} to compute $I$ for the $SU(3)$ bundle since it is not
$\t$-invariant. In fact, Equation~\eqref{Itau} predicts
$I_{SU(3)}=60$, which is wrong, but this is not a problem since
Equation~\eqref{Itau} is only valid for $\t$-invariant bundles anyway.

In our example, the hidden sector group was completely broken, which may not
be phenomenologically desirable.
However, we could also consider models with unbroken hidden sector groups, by
adding points to $\nabla$ over the ray $R_2$, as in the six dimensional
situation~\cite{\CF}. This leads to a large number of
possibilities for models with three generations. Since we have
more than one choice for $\eta$ and $\g$ yielding three generation models
to begin
with, we see that we can construct many \cyf s that yield three generation
models. Note also that while we have only done the analysis for $E_6$, a
similar analysis can also be done for the GUT groups $SO(10)$ and $SU(5)$.

In passing, we note that the methods of Klemm
{\it et al\/}.~\Ref\KLRY{A.~Klemm, B.~Lian, S-S.~Roan and S-T.~Yau,
Nucl. Phys. {\bf B497} (1997) 173, hep-th/9609239.}
can be used to
identify all the divisors with arithmetic genus 1, \ie\ those that can
contribute to the superpotential. We list those divisors below: 
$$\eqalign{&{
\hbox{\tt(0,0,0,2,3),(0,0,1,0,0),(0,0,1,0,1),(0,0,1,2,3),}}\cr
&{\hbox{\tt(0,0,2,1,1),(0,0,2,1,2),(0,0,2,2,3),(0,0,3,2,3),(0,1,6,2,3)}}\cr}$$
Note that the first of these is a horizontal divisor, and so does not
contribute to the superpotential in F-theory.
\newpage
\section{fivebranes}{Fivebranes and Extremal Transitions}
In the work of Friedman, Morgan and Witten~\cite{\FMW},
as well as in all the examples
studied so far, one only considers models with
$\eta(V_1)+\eta(V_2)=12c_1(B_2)$. Among other things, this ensures that all the
fivebranes on the heterotic side are wrapping the elliptic fibre of the \cyt .
We can now ask what happens when the $\eta(V_i)$ do not add up to $12c_1(B_2)$,
say, $12c_1(B_2)-\eta(V_1)-\eta(V_2)=C$, where $C$ is a divisor in $B_2$.
By Equation~\eqref{hetanomaly},
the cohomology class of the fivebranes includes the class $C\sigma$.
One necessary constraint on the class $C$ is that it must
correspond to an effective divisor, a condition obtained
in~\cite{\DLOW}. The authors of that work also study the
moduli space of these fivebranes. Here, we will discuss the F-theory picture
dual to this situation.

The natural interpretation of these fivebranes (which was also proposed
by~\cite{\DLOW}) is that they wrap holomorphic curves in the base whose class
is precisely
$C$. For the purposes of this paper, we will only consider the situation when
the curve $C$ is a $\IP^1$.
We now propose a heuristic argument for guessing the F-theory dual to
this picture. Later, we will construct models that support our argument. In the
absence of fivebranes wrapping curves in $B_2$,
the base of the \cyf\ on the F-theory side is a $\IP^1$ fibred over $B_2$.
Thus for any $\IP^1$ in the heterotic $B_2$, we have a corresponding $\IP^1$
in the F-theory $B_3$. Consider the limit when this $\IP^1$
becomes large, but the rest of the \cyf\ remains small. In this limit, we
arrive at something that looks like a six dimensional situation. Wrapping a
fivebrane on the large $\IP^1$  then looks like turning on a fivebrane in the
six dimensional sense, which corresponds to a blow up of the F-theory base in
six dimensions. Now, the F-theory base $B_3$ can be regarded as being
fibred over the wrapped $\IP^1$, so we are really blowing up over this wrapped
$\IP^1$. Thus we arrive at our F-theory dual: wrapping a fivebrane over a
$\IP^1$ in the heterotic base corresponds to blowing up the corresponding 
$\IP^1$ in $B_3$ into a ruled surface (a $\IP^1$ fibred over the
original $\IP^1$).
This is therefore a ``fibred'' version of the six dimensional story when we
have extra tensor multiplets. Carrying this analogy further, we note that the
extremal transition on the F-theory side corresponding to the appearance of an
extra fivebrane (in six dimensions) was described in~\cite{\rMVII}. In that
work, the
authors show that the appearance of an extra fivebrane corresponds to
replacing a singular point by the del Pezzo surface $dP_8$. Similarly, one can
argue
that in the four dimensional case, wrapping a fivebrane around a $\IP^1$ in
$B_2$ corresponds to replacing the corresponding (singular) curve in the
\cyf\ by a $dP_8$ fibred over it.

\noindent$\underline{\it 5.1}\quad$
Let us consider some specific examples to illustrate this idea. Consider
F-theory compactified on the \cyf\ which is elliptically fibred over 
$\IF_{100}$ (we refer the reader Table~\tabref{fmnp} and
Equation~\eqref{fmnprels}, where now $n=p=0$).
The heterotic dual consists of an $E_8\times E_8$ bundle over the
\cyt\ which is elliptically fibred over $\IF_1$, with $\eta=6c_1(\IF_1)$ for
each of the two $E_8$ bundles. The Hodge numbers of the \cyf\ are
$h_{11}=4, h_{31}=2916, h_{21}=0, h_{22}=11724$ and the Euler number is
$\chi=17568$. The cohomology class of the fivebranes is then given by
Equation~\eqref{hetanomaly} to be $c_2(\IF_1) + 91c_1^2(\IF_1)$ (hence they all
wrap the elliptic fibre), and when integrated over $B_2$ gives
$n_5=732=n_3={\chi\over24}$, as expected. Also, using Equations~\eqref{chernE}
and~\eqref{Itau} to count the moduli of the bundles, we find that each $E_8$
has $I=1336$, so that $1+h_{21}(Z)+I_1 +I_2 + h_{21}(X) = 1 + 243 + 1336 +
1336 + 0 = 2916 = h_{31}(X)$, as expected.

\noindent$\underline{\it 5.2}\quad$
Consider now the situation when the second $E_8$
bundle has $\eta=6c_1(\IF_1)-C_0$ instead. Then the cohomology class of the
fivebranes is $C_0\sigma + c_2(\IF_1) + 91c_1^2(\IF_1) + 15 C_0^2 -
45 C_0.c_1(\IF_1)$.
We interpret the first term as a fivebrane
wrapping the zero section of $\IF_1$ (in what follows, we will use the same
notation for the cohomology class of a divisor in the base, and the
corresponding curve in the base, and let the context distinguish between the
two), and the rest as fivebranes wrapping the
elliptic fibre of $Z$. Integrating over $\IF_1$, we find that there are
$n_5^{\ca E}=672$ fivebranes wrapping the elliptic fibre. 
On the F-theory side, the new \cyf\
is determined by the change in $\eta(V_2)$ --- we find that the only way to
achieve
this is to blowup $\IF_{100}$ by adding the ray $(0,1,-1)$. Under the natural
projection of the base to $\IF_1$, this new ray projects to $C_0$.
Furthermore, this ray subdivides the two dimensional cone generated by
$(0,1,0)$ and $(0,0,-1)$, and therefore corresponds to a blowup of a curve
into a rational surface, and not a blowup of a point into
a $\IP^2$.
Hence we conclude that we are really blowing up
the curve $C_0$ into a rational surface. It follows that the extremal
transition from the original \cyf\ to the new \cyf\ is indeed described by
replacing a curve in the original fourfold by a $dP_8$ fibred over
it\Footnote{I am grateful to Albrecht Klemm for explaining this point.}.

The Hodge numbers of this fourfold are $h_{11}=5, h_{31}=2675, h_{21}=0,
h_{22}=10764$ and the Euler number is $\chi=16128$. From the tadpole
anomaly cancellation condition~\eqref{tadpole} we expect
$n_3={\chi\over24}=672$ threebranes, which is precisely $n_5^{\ca E}$, the
number of fivebranes wrapping the elliptic fibre. We see therefore that the
fourfold distinguishes between fivebranes wrapping the elliptic fibre and
fivebranes wrapping curves in the base. The first contribute to the number of
threebranes, and hence to
the Euler number of the fourfold, while the second are
seen as blowups in the base $B_3$ and hence contribute to $h_{11}$. Note that
$h_{11}$ agrees with Equation~\eqref{h11new}, since blowing up $\IF_{100}$
once increases its $h_{11}$ to $4$, and hence increases $h_{11}(X)$
to $5$. Finally, using Equations~\eqref{chernE}
and~\eqref{Itau} to count the moduli of the bundles, we find that the second
$E_8$ now has $I=1095$, so that $1+h_{21}(Z)+I_1 +I_2 + h_{21}(X) = 1 + 243 +
1336 + 1095 + 0 = 2675 = h_{31}(X)$, as expected.

\noindent$\underline{\it 5.3}\quad$
Now, in the above example, instead of wrapping the fivebrane around $C_0$, we
can choose to wrap it around say, $f$, by choosing the second $E_8$ bundle to
have $\eta = 6c_1(\IF_1) - f$. The cohomology class of the fivebranes is then
$f\sigma + c_2(\IF_1) + 91c_1^2(\IF_1) - 45 f.c_1(\IF_1)$. The first term
then corresponds to a fivebrane wrapping the fibre of $\IF_1$, while the rest
describe fivebranes wrapping the elliptic fibre of $Z$. Integrating over
$\IF_1$, we find that there are $n_5^{\ca E}=642$ fivebranes wrapping the
elliptic fibre. On the F-theory side, the new \cyf\
is determined by the change in $\eta(V_2)$ --- which we achieve
by blowing up $\IF_{100}$ by adding the ray $(-1,0,-1)$. Under the natural
projection of the base to $\IF_1$, this new ray projects to $f$.
Furthermore, this ray subdivides the two dimensional cone generated by
$(-1,0,0)$ and $(0,0,-1)$, and therefore corresponds to a blowup of a curve
into a rational surface, and not a blowup of a point into
a $\IP^2$.
Again, we conclude that we are really blowing up
the curve $f$ into a rational surface. It follows also that the extremal
transition from the original \cyf\ to the new \cyf\ is again described by
replacing a curve in the original fourfold by a $dP_8$ fibred over
it.

The Hodge numbers of this fourfold are $h_{11}=5, h_{31}=2555, h_{21}=0,
h_{22}=10284$ and the Euler number is $\chi=15408$. From the tadpole
anomaly cancellation condition~\eqref{tadpole} we expect
$n_3={\chi\over24}=642$ threebranes, which is precisely $n_5^{\ca E}$, the
number of fivebranes wrapping the elliptic fibre. We see again that the
fourfold distinguishes between fivebranes wrapping the elliptic fibre and
fivebranes wrapping curves in the base. As in the previous example,
the first contribute to the number of
threebranes, and hence the Euler number of the fourfold, while the second are
seen as blowups in the base $B_3$ and hence contribute to $h_{11}$.

Finally,
using Equations~\eqref{chernE} and~\eqref{Itau} to count the moduli of the
bundles, we find that the second $E_8$ now has $I=974$, so that
$1+h_{21}(Z)+I_1 +I_2 + h_{21}(X) = 1 + 243 +
1336 + 974 + 0 = 2554$, which is one less than $h_{31}(X)$. The missing modulus
must be interpreted as a deformation of the curve $f$ that is
being wrapped. Let us verify that such is indeed the case. We note that the
heterotic base $\IF_1$ can be viewed as a blowup (by a $\IP^1$)
of $\IP^2$ at a point $p$. This blowup is in fact a $-1$ curve, and is
precisely the zero section $C_0$. Note that since $C_0$ has negative
self-intersection, it cannot be deformed --- thus there are no additional
complex structure moduli when a fivebrane wraps $C_0$, which explains the 
matching of the bundle moduli and $h_{31}(X)$ in example {\it 5.2\/}.
However, the curve in the class $f$ can be deformed (since $f^2=0$). In fact,
it is a $\IP^1$ curve in $\IP^2$ which passes
through the point $p$ which is being blownup, since $f.C_0=1$. Now, a general
$\IP^1$ curve in $\IP^2$ can be described by the equation $ax + by+ cz=0$,
where $x,y$ and $z$ are homogeneous coordinates on $\IP^2$. This has two
deformation parameters, since the equation for the curve is invariant
under a rescaling of $x,y$ and $z$. However, requiring the curve to pass
through a
given point reduces the number of complex deformations by one and so
we have only one parameter. We claim that this parameter is precisely the
missing modulus in $h_{31}(X)$. This also implies that the second of
Equations~\eqref{bundlehodge} must be modified to include a term corresponding
to the deformation moduli of the curves being wrapped.

\noindent$\underline{\it 5.4}\quad$
Let us consider now the situation when the
class of the fivebranes wrapping curves in the base is $C_0+f=C_\infty$,
by setting, say, $\eta(V_2) = 6c_1(\IF_1) - C_0 - f$. The
cohomology class of the fivebranes is now
$(C_0+f)\sigma + c_2(\IF_1) + 91c_1^2(\IF_1) + 15 (C_0+f)^2
- 45 (C_0+f).c_1(\IF_1)$. Integrating over $\IF_1$, we obtain
$n_5^{\ca E}=612$ fivebranes wrapping the elliptic fibre. However,
on the F-theory side, we now have two possibilities: we can either blowup once
over $C_\infty$ by adding the ray $(0,-1,-1)$, or we can blow up once over
$C_0$ and once over $f$ by adding the rays $(0,1,-1)$ and $(-1,0,-1)$. 
We interpret the first case as corresponding to a fivebrane wrapping
$C_\infty$, and the second as corresponding to two fivebranes: one wrapping
$C_0$, and the other wrapping $f$. In both instances, we are subdividing two
dimensional cones, so that we are blowing up curves into ruled surfaces.

However, the Hodge numbers of the corresponding \cyf s are different.
The first fourfold has Hodge numbers $h_{11}=5, h_{31}=2435, h_{21}=0,
h_{22}=9804$ and Euler number $\chi=14688$, while the second has $h_{11}=6,
h_{31}=2434, h_{21}=0, h_{22}=9804$ and Euler number $\chi=14688$. 
Note, however, that the bundle data (the $\eta_i$) are identical, and the class
of the fivebranes is also identical. Furthermore, the tadpole
anomaly cancellation condition~\eqref{tadpole} predicts
$n_3={\chi\over24}=612$ threebranes, in both cases, which is precisely
$n_5^{\ca E}$, the number of fivebranes wrapping the elliptic fibre.
The only difference is the specific
curve in $B_2$ that is being wrapped by the fivebranes - in one case we
have a single fivebrane wrapping the infinity section leading to one blowup in
the base of the fourfold dual, while in the other, we
have two fivebranes, one wrapping the fibre, and the other wrapping the
zero section of $\IF_1$, leading to two blowups in the base of the dual
fourfold. Since these two curves lie in the same class, it is
conceivable that there is a degeneration that takes one curve to the other.
If we were able to follow this degeneration, then we would obtain a heterotic
version of the extremal transition on the F-theory side, where we replace
one $dP_8$ fibred over $C_\infty$ by two $dP_8$'s, one fibred over $C_0$, and
the other over $f$.

Finally,
using Equations~\eqref{chernE} and~\eqref{Itau} to count the moduli of the
bundles, we find that the second $E_8$ now has $I=853$, so that
$1+h_{21}(Z)+I_1 +I_2 + h_{21}(X) = 1 + 243 +
1336 + 853 + 0 = 2433$, which is two less than $h_{31}(X)$ for the first case
and one less than $h_{31}(X)$ for the second. Recall that the second case has
one fivebrane wrapping the zero section $C_0$, and one wrapping $f$. From our
previous example, we know that the curve $f$ has one deformation
parameter, while $C_0$ has none, which accounts for the discrepancy in this
case. The two missing moduli in the first case, when the fivebrane wraps
$C_\infty$, can also be interpreted as deformation parameters of the curve
that is being wrapped. To see this, recall our discussion from
example {\it 5.3\/}, where we regard $\IF_1$ as a blowup of $\IP^2$ at a
point $p$, with the exceptional
$\IP^1$ identified with $C_0$. Now, $C_\infty^2=1$, so we see that it
can indeed be deformed. Furthermore, $C_\infty.C_0=0$, so we see that it is
in fact a curve in the $\IP^2$ that does not pass through $p$. Such a curve
is then a generic curve in $\IP^2$ with two moduli, as discussed above. These
are precisely the additional moduli needed to account for the observed
discrepancy in $h_{31}(X)$. Finally, note that the degeneration of $C_\infty$
into two curves is easy to describe in this picture. When the curve $C_\infty$
is deformed so as to pass through the point $p$ in $\IP^2$, it splits up into
two curves, one corresponding to $f$, and the other to the exceptional divisor,
which is precisely $C_0$. In the process, we lose exactly one deformation
modulus, so $h_{31}(X)$ drops by one. Also, since one curve ($C_\infty$)
now splits in two ($C_0$ and $f$), we gain one K\"ahler modulus, so
$h_{11}(X)$ increases by one. Thus we have described the topology changing
extremal transition from the first \cyf\ to the second by a degeneration
of curves on the heterotic side. Note, however, that this discussion has been
purely classical. We would expect this picture to be modified by quantum
corrections, in particular, the effects of the superpotential.

More generally, if the heterotic base is $\IF_m$, then $C_\infty=C_0+mf$, and
each fivebrane in this class corresponds either to a single blowup (over
$C_\infty$) of the \cyf\ base, or $m+1$ blowups, $m$ over $f$ and one over
$C_0$. The choice of the curves being wrapped will then give fourfolds with
different Hodge numbers but the same Euler number, since the number of
fivebranes wrapping the elliptic fibre of $Z$, and hence the number of
threebranes, is the same. Degenerations from one such curve in $\IF_m$ to the
other could then be used to follow the topology changing
extremal transition from one \cyf\ to the other. 
A detailed description of these transitions is currently under
investigation. We hope to report on this in the future. 

\newpage
\section{fin}{Discussion}
In this paper, we have studied F-theory compactifications on \cyf s dual to
heterotic compactifications on \cyt s with some choice of vector bundle. We
have shown how toric data (in particular, the polyhedron $\nabla$) for
the \cyf\ encode the information about the heterotic dual, in particular, the
$\eta$ class of the heterotic vector bundle. We have found that reflexivity of
the fourfold polyhedron imposes a restriction on $\eta$, which is related to
the stability of the corresponding vector bundle. We have also discussed the
construction of \cyf s corresponding to three generation models. We note that
this construction requires specifying the value of the four flux in addition
to the \cyf . While the four flux is, in general, independent of the \cyf ,
we have seen that requiring the number of threebranes to be non-negative puts
bounds on its value. Thus, not all \cyf s can correspond to three generation
models.

We have also explored the issue of heterotic fivebranes wrapping curves in
the base. We find that these correspond to blowups in the base of the \cyf .
We have also seen that if the cohomology class of the fivebranes corresponds
to a reducible divisor in the heterotic base, then there is an ambiguity in
choosing the curves
that are being wrapped. This corresponds on the F-theory side to different
numbers of blowups, and thus topologically different fourfolds with different
Hodge numbers (but same Euler number), and the extremal transitions between
these fourfolds can be described (classically) in terms of degenerations of
curves in the base of the heterotic threefold.

Thus, we see that, as in the six dimensional
situation~\cite{\CF -\CPRmat}, toric geometry provides a natural
arena for discussing F-theory/heterotic duality in four dimensions. The map
between toric data and vector bundle data presented in this paper allows us
to construct a very rich class of dual pairs. There are, however, many issues
in heterotic/F-theory duality which we
have not addressed at all in this paper. We have, for instance, not
attempted to identify the origin of the matter (in particular, the chiral
matter) in the fourfold.  Also, we do not have an understanding of the effects
of quantum corrections on the proposed duality. Further, we have
not studied models with
non-perturbative (from the heterotic perspective) gauge groups, nor have
we addressed the question of mirror symmetry in fourfolds and its heterotic
interpretation. Clearly, we have merely scratched the surface of this subject,
and there are many interesting issues to be examined.
\bigskip

{\centerline {\bf Acknowledgements}}
\noindent{I am grateful to P.~Candelas, G.~Curio, D-E.~Diaconescu,
R.~Friedman, A.~Klemm, P.~Mayr, B.~Ovrut, E.~Perevalov, S.~Sethi, D.~Waldram
and E.~Witten for useful discussions during
various stages of this work. This work was supported in part by
NSF Grant Math/Phys DMS-9627351.}
\bigskip
{\centerline {\bf Note Added}}
\noindent Although the argument in \SS{4.3} leading to Eqn.~\eqref{etamin} is
correct, it does not yield
the sharpest lower bound that can be imposed on the $\eta$ class, since it
only uses the order of vanishing of the discriminant locus. It
is in fact possible to obtain a stricter bound
by using Tate's algorithm~\cite{\Tate} in full
generality~\ref{P.~Berglund and P.~Mayr, hep-th/9904114.}. 
\bigskip

\immediate\closeout\referencewrite\referenceopenfalse
\line{\bf\hfil References\hfil}\bigskip\parindent=0pt\input referenc.texauxil
\end